\documentclass[nonacm,sigconf]{acmart}
\usepackage{amsmath}
\usepackage{ulem}
\usepackage{subfigure}
\usepackage{algorithm2e}
\usepackage{algorithmic}
\usepackage{multirow}
\usepackage{makecell}
\usepackage{tabularx}
\usepackage{multicol}
\usepackage{nicematrix}
\usepackage{array}
\usepackage{tabularray}
\usepackage{enumitem}
\setlist[itemize]{leftmargin=*}
\usepackage{adjustbox}
\usepackage{marginnote}
\usepackage{titlesec}
\usepackage{amsthm}
\setcopyright{none}
\newcommand{\rjnote}{\color{blue}}
\newcommand{\wqnote}{\color{purple}}
\newcommand{\issues}{\color{brown}}
\definecolor{ao(english)}{rgb}{0.0, 0.5, 0.0}
\newcommand{\cwnote}[1]{{\color{ao(english)}#1}}
\titleformat{\paragraph}
{\normalfont\normalsize\bfseries}{\theparagraph}{1em}{}
\titlespacing*{\paragraph}
{0pt}{0.25ex plus 1ex minus .2ex}{0.5ex plus .2ex}
\usepackage{relsize}
\newcommand*{\defeq}{\stackrel{\mathsmaller{\mathsf{def}}}{=}}
\newcolumntype{Y}{>{\hsize=\hsize\centering\arraybackslash}X}
\newcolumntype{Z}{>{\hsize=1.5\hsize\centering\arraybackslash}X}

\begin{document}

\title{Simulation-based Validation for Autonomous Driving Systems}

\author{Changwen Li}
\affiliation{%
  \institution{State Key Laboratory of Computer Science, ISCAS}
  \city{Beijing}
  \country{China}
}
\author{Joseph Sifakis}
\affiliation{%
  \institution{Univ. Grenoble Alpes, CNRS, Grenoble INP, VERIMAG}
  \city{Grenoble}
  \country{France}
}
\author{Qiang Wang}
\affiliation{%
  \institution{Academy of Military Sciences}
  \city{Beijing}
  \country{China}
}
\author{Rongjie Yan, Jian Zhang}
\affiliation{%
  \institution{State Key Laboratory of Computer Science, ISCAS}
  \city{Beijing}
  \country{China}
}


\begin{abstract}
Simulation is essential to validate autonomous driving systems. However, a simple simulation, even for an extremely high number of simulated miles or hours, is not sufficient. We need well-founded criteria showing that simulation does indeed cover a large fraction of the relevant real-world situations. In addition, the validation must concern not only incidents, but also the detection of any type of potentially dangerous situation, such as traffic violations.

We investigate a rigorous simulation and testing-based validation method for autonomous driving systems that integrates an existing industrial simulator and a formally defined testing environment. The environment includes a scenario generator that drives the simulation process and a monitor that checks at runtime the observed behavior of the system against a set of system properties to be validated. 
The validation method consists in extracting from the simulator a semantic model of the simulated system including a metric graph, which is a mathematical model of the environment in which the vehicles of the system evolve. 
The monitor can verify properties formalized in a first-order linear temporal logic and provide diagnostics explaining their non satisfaction.
Instead of exploring the system behavior randomly as many simulators do, we propose a method to systematically generate sets of scenarios that cover potentially risky situations, especially for different types of junctions where specific traffic rules must be respected. 
We show that the systematic exploration of risky situations has uncovered many flaws in the real simulator that would have been very difficult to discover by a random exploration process. 
\end{abstract}


\maketitle

\section{Introduction}
Autonomous driving systems (ADS) are real-time distributed systems involving components with partial knowledge of their environment, pursuing specific goals while the collective behavior must meet given global properties. They are probably the most difficult systems to design and validate, as they are built from heterogeneous components subject to temporal and spatial dynamism. They operate in unpredictable environments whose topological and geometric properties constrain the behavior of their agents. 
These characteristics challenge the application of rigorous model-based development and validation techniques that we have successfully applied to critical systems such as aircraft systems \cite{Basu2011RigorousCS}. In particular, formal methods are defeated by the complexity of these systems and can be applied only to their components \cite{sifakis-framework}. 

In the face of these problems, global validation through simulation and testing appears to be a viable approach to obtain evidence of the trustworthiness of ADS. Industrial players have developed powerful simulation environments and are reporting impressive numbers of simulated miles as evidence that their products are sufficiently safe 
\footnote{Waymo: “Off road, but not offline: How simulation helps advance our waymo driver,” https://blog.waymo.com/2020/04/off-road-but-not-offline--simulation27.html}. 
However, this argument is not tenable, for the reason that not all simulated miles are equally effective. It is necessary to explain how simulated miles relate to "real miles". This requires a deeper understanding, through rigorous modeling, of the extent to which all relevant system configurations are explored. 

Currently industrial simulators, which are mostly built on top of game platforms, favor a certain realism of modeling but they lack semantic awareness required for rigorous validation. Semantic awareness implies the possibility to extract from the simulated system software, a semantic model with a well-defined notion of system state as the distribution of vehicles on a map with all their kinetic and time attributes. Such a model should define an abstract system behavior with a notion of execution step and should respect fundamental properties of time and space. For example, if there are two different runs leading from one system state to another, the travel time and distance along these 
runs must be identical. 
An adequate semantic model is important for the definition of execution runs and predicates on system configurations needed for system validation. 

Simulators should also support multiscale/multigrain modeling. Validation aims at checking that the system and its vehicles satisfy three different types of goals, each requiring modeling at different level of detail and time scale. 
Short-term goals are to keep the system out of dangerous states and to meet strict real-time safety constraints. They would aim to avoid collisions by keeping distance from obstacles within certain limits or by following a pre-defined trajectory. Their validation requires detailed models of the vehicle’s static environment and of the obstacles around it, built in real time from sensory information. This information can be complemented by pre-stored knowledge from a repository, e.g.,
detailed maps, characteristics of types of obstacles for better goal management and adaptation. 

Medium-term goals concern the transition between predefined operating modes under specific time constraints and system reconfiguration adapting to dynamically changing situations. These goals require performing maneuvers such as overtaking or crossing junctions of various types. 
Their validation requires a fairly detailed model of the vehicle's external environment combining real-time sensory information with pre-recorded information beyond the vehicle's visibility horizon to anticipate maneuvers.

Long-term goals intend to satisfy various types of non-critical properties such as completing a trip by reaching a destination, or optimizing fuel consumption. Their validation requires high-level map models of the vehicle’s route adequately enriched with semantic information about traffic performance, rerouting, etc. 


Ideally, validation environments for ADS integrate three collaborating tools, that is 1) a Simulator; 2) a Scenario Generator; and 3) a Monitor. 

The Simulator executes an ADS model obtained as the composition of two entities. 
The first entity is a model of the environment, which is usually represented by a kind of map with all relevant information to determine the system state. The information includes the positions of the vehicles and their kinetic attributes, the obstacles around each vehicle as well as signaling information used to enforce traffic rules. 
The second entity involves behavioral models of the vehicles and their possible interactions. The Simulator exhibits cyclic behavior alternating between concurrent executions of the vehicle models for a given lapse of time and the computation of the resulting system state on the environment model.

The Scenario Generator provides test cases that drive the simulated model and interacts with the Simulator in an adaptive manner. For a given system state, it generates commands executed by the vehicles so as to bring the system in configurations of interest. Scenarios apply strategies that dynamically change the applied commands based on the observed system reactions. They can modify long and mid-term goals such as routes and performing maneuvers while short-term goals are left to the vehicle control system. 

The Monitor plays the role of an Oracle in testing terminology and checks that the system behavior is compliant with given properties. It observes the system behavior as sequences of global states and applies a runtime verification method of the properties. These can be specific properties or general properties such as traffic rules.

Currently, there is still a big gap between the state-of-the-art and the needs for validation of ADS. Many building blocks are missing for a rigorous and effective application of the above principle.  This work attempts a quick assessment of existing results and discusses how the functionality of existing simulation tools can be enhanced by connecting them to rigorous validation tools.

The main idea is to connect an industrial simulator developed in an ad hoc manner, to validation tools through an API exporting a semantic model of the simulated system behavior. Defining such a model involves a few technical difficulties in particular because it should use some symbolic representation of the system state, which is essential for rigorous analysis. We show how for the considered simulator it is possible to define such a semantic model based on maps defined as metric graphs. The semantic model is implemented using an API through which the relevant information needed by the validation tools is exported.
Based on the defined semantic model, we develop results in two directions. 

The first concerns the runtime verification of properties described in a linear temporal logic built from atomic propositions that characterize the state of the system's semantic model. These propositions concern, on one hand,  topologic and geometric properties of a map, and on the other hand, relations between the vehicles evolving in the system and their surrounding obstacles. We show how the proposed logic can be used to specify system properties and how formulas can be validated using existing runtime verification techniques.
%


 The second direction is the generation of scenarios and their actual application to control the system behavior. Scenario generation is guided by coverage criteria of risky situations in junctions that can lead to traffic violations and potentially to accidents. We show how high-risk scenario generation can uncover flaws in a real simulator that would have been very difficult to detect through random exploration. 

Finally, we provide a review of related work that demonstrates the novelty of our approach and indicates avenues for the future development.


\section{An overview of the validation framework}

\begin{figure}[h]
\centering
\includegraphics[width=0.48\textwidth]{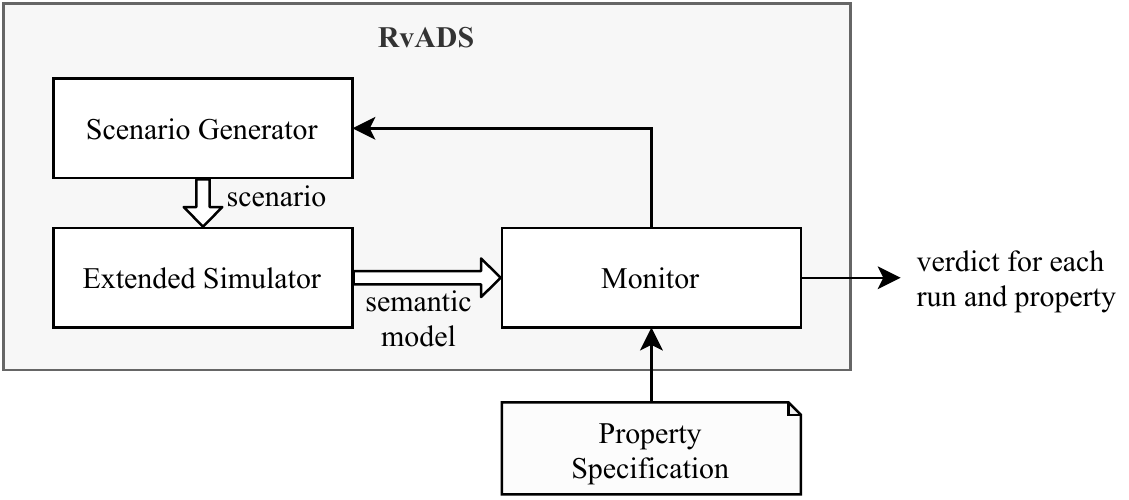}
\caption{\label{fig:architecture} An overview of the proposed RvADS framework}
\end{figure}

Fig.~\ref{fig:architecture} presents an overview of the proposed RvADS framework,
 which integrates the LGSVL autonomous driving Simulator~\cite{rong2020lgsvl} with a Scenario Generator and a Monitor
 for the validation of ADS (Fig.~\ref{fig:architecture}). 
We thus reuse a publicly available simulator  
 that allows for realistic system modeling,
 in a validation environment for on-line checking of ADS against given properties.

 The Scenario Generator drives the simulation process. 
 It generates scenarios as the coordinated sequences of actions 
 executed by the agents in the Simulator. 
 Thus, scenarios play the role of test cases intended 
 to drive the simulated system towards specific configurations, 
 for example to explore particular high-risk configurations or to meet specific coverage criteria.
 
 The Simulator runs the system model that consists of both the behavior models of traffic agents 
 (e.g. mobile agents such as vehicles and pedestrians, 
 and non-moving objects such as traffic signals) 
 and the model of the static environment. The latter includes the map and all relevant information regarding the system state. 
 The state of traffic agents consists of their kinematic attributes 
 and their positions on the map. 
  The state of the system model is defined by the positions of the agents and objects in their
environment with their states. It determines the possible interactions
between agents, objects and their physical environment.
The resulting global behavior should satisfy safety and performance
properties, in particular these implied by traffic rules.
 The Simulator has a periodic behavior. In each cycle, it provides each agent with
 information about the free space in its vicinity and directives to
 execute actions from the Scenario Generator. 
 The agents, after moving for a period in their respective free
 space, report their movements to the Simulator. 
 The latter updates the positions of the agents on the map before initiating a new cycle.

To link the simulation and the validation process, we have extended the Simulator. 
Particularly in order to explore specific situations, we modified the agents' controllers to follow predefined scenarios. 
A second modification is the development of an API to export an abstract semantic model of the simulated system on which the Monitor can check the properties to be validated.


The Monitor checks whether the runs of the simulated system
 satisfy given properties expressed as logical formulas.
 It evaluates the atomic propositions of the formulas based on the semantic model exported through the Simulator API. 
 It applies runtime verification techniques to detect property violations 
 from the observed behavior of the system. 
 In addition, it can provide diagnostics and key performance indicators. 
 We assume that the specified properties express: 
 1) either safety, i.e., for any execution, all system states satisfy a certain safety constraint; 
 2) or limited reachability, i.e., for any execution, a desirable condition will be satisfied within a given number of steps.

\section{State of the practice in ADS simulation}

Simulators are essential for testing and evaluating ADS. 
 Popular industrial simulators such as CARLA \cite{Dosovitskiy17} 
 and LGSVL \cite{2020arXiv200503778R} not only provide the 3D rendering of environment, 
 but also support the modeling and testing of fundamental ADS functionalities 
 such as perception, path planning, and vehicle control.

CARLA is an open-source autonomous driving simulator 
 built on top of the Unreal Engine 4 game engine 
 and using the OpenDRIVE standard to define roads and urban settings.
 A simulation in CARLA combines two components: 
 (i) the CARLA server that computes the physics and renders the scene;
 (ii) the client scripts in Python or C++ API implementing the perception, 
 planning and vehicle control functionalities.
 The control parameters (e.g., throttle, brake and steering) 
 computed by the client scripts are then sent back to the CARLA server, 
 thus forming a client-server interaction loop.
 During the simulation process, 
 CARLA also provides client functions to extract a vehicle's current state 
 (e.g. location, velocity, and acceleration) and 
 supports exporting a sequence of states in simulation for replays and further post-inspections.

LGSVL is another open platform for autonomous driving simulation.
 It is equipped with an API to export the current status of the vehicles 
 (3D position, speed, orientation, and angular velocity) and eventually control them. 
 In addition, it can use the autonomous driving software of the Apollo or Autoware platform 
 to control the simulated vehicles. 
 Finally, LGSVL allows the use of HD maps in some common formats,
 for example OpenDrive and Apollo Map.

It should be noted that the above-mentioned simulators do not
 provide all the information needed for a rigorous validation.
 Although LGSVL is capable of exporting HD maps with geometric
 information, it does not provide the semantic information 
 essential for checking traffic rules, e.g., junction types and
 associated signage. Furthermore, the exported reports provide
 absolute 3D positions of the agents, while important information
 about their relative positions on the map is missing. Therefore, the
 link between the state of a vehicle and its position on the map is
 not explicit and cannot be found without additional computation. However, such a computation is  needed as property validation requires relative positions. Finally, LGSVL does not allow  the specification of
 vehicle itineraries and the control of their execution, which limits the testing capabilities of the
 scenarios, e.g., to achieve test objectives characterizing specific
 system configurations.

\section{Connecting simulation to validation}

This section presents the construction of the semantic model 
 built from the information extracted from the LGSVL simulator.   
 To facilitate the discussion, we make the following assumptions:
\begin{itemize}
    \item The map model is a customized version of the OpenDRIVE specification. 
    \item All the roads are single-lane, and equipped with the necessary signals to enforce the traffic rules.  
    \item For the lanes, we ignore their width and simply represent them by their center lines. 
\end{itemize}

\subsection{Map model formalization}

The basis of the semantic model is a metric graph \cite{sifakis-framework}, 
 defining the set of positions on which the simulated vehicles are
 located. We first present the HD map model used by the LGSVL
 simulator and then the transformation into a formal metric graph model.


The maps adopted in the state-of-the-art simulators are usually high-definition and have various map annotations. 
 For example, the annotations of the map model in the LGSVL simulator involve lanes,
 junctions, self-reversing lanes, and other features. 
 The annotations of lanes include waypoints and boundary lines. 
 The annotations of intersections include lanes, traffic signals, traffic signs, and stop lines. 
 The annotations of self-reversing lanes have elements under traffic lanes and intersections. 
 The annotations of other features involve pedestrian paths, crosswalk, parking space, etc.   
 Among these annotations, lanes are the fundamental elements to build roads and junctions.


 In this paper, we consider single lane roads and junctions. 
 Nevertheless, the approach followed can be extended to multi-lane roads without significant changes for the problems studied. 

 A lane is denoted by $\ell=(l_1,\tau,\lambda,l_2)$, 
 where $l_1$ and $l_2$ record the locations of the endpoints of $\ell$ 
 and the direction of the lane is from $l_1$ to $l_2$, 
 $\tau$ represents the turn type of the lane
 (such as \texttt{left}, \texttt{right}, \texttt{straight}), 
 and $\lambda$ is the label type of the lane (i.e., either in a road or in a junction). 
 For example, in the sub-map of Fig.~\ref{fig:map}, 
 there is a road with lane $\ell_1$ and a junction with two lanes $\ell_2$ and $\ell_3$, 
 positioned with respect to their center-lines. 
 The lane $\ell_1$ in the road is denoted by $(l_2,\texttt{straight}, \texttt{road},l_1)$, 
 and  the lane $\ell_3$ in the junction is denoted by $(l_4,\texttt{left},\texttt{junction},l_2)$.

 \subsubsection{Metric graph.}

 We use metric graphs~\cite{sifakis-framework} for the
 formalization of the map model provided by the Simulator. These
 are directed graphs whose edges are labeled with segments that
 carry geometric information characterizing the road structure
 between their endpoints. 


A metric graph is formally defined by a tuple $G=(V, \mathcal{S}, E)$, 
 where $V$ is a finite set of vertices, $\mathcal{S}$ is a set of segments with a partial concatenation operator $\cdot:\mathcal{S}\times \mathcal{S}\rightarrow \mathcal{S}\cup \{\bot\}$, 
 and $E \subseteq V\times\mathcal{S}^+\times V$ is a set of edges labeled 
 by  segments of non-zero length (denoted by $\mathcal{S}^+$).
 An edge $e=(v,\mathbf{s},v')$ is denoted by $v\xrightarrow[]{\mathbf{s}}v'$ for brevity.
 For example, the metric graph representing the sub-map of Fig.~\ref{fig:map} is shown in Fig.~\ref{fig:metricgraph}. 
There are three vertices in the
graph and three edges labeled with segments $\mathbf{s}_1$, $\mathbf{s}_2$, $\mathbf{s}_3$ corresponding to the three lanes of the map. 


In addition to the vertices,  we introduce the concept of \textit{positions} on a segment $\mathbf{s}$ as follows:
\begin{itemize}
    \item $\mathbf{s}(-,\eta)$ and $\mathbf{s}(\eta,-)$ are the positions of distance $\eta$ from the beginning and the end of $\mathbf{s}$, respectively. 
    \item $\mathbf{s}[-,\eta]$ and $\mathbf{s}[\eta,-]$ are the segments defined respectively between the beginning of $\mathbf{s}$ and $\mathbf{s}(-,\eta)$, and between $\mathbf{s}(\eta,-)$ and the end of $\mathbf{s}$.
\end{itemize}

\begin{figure}[]
\centering
\includegraphics[width=0.35\textwidth]{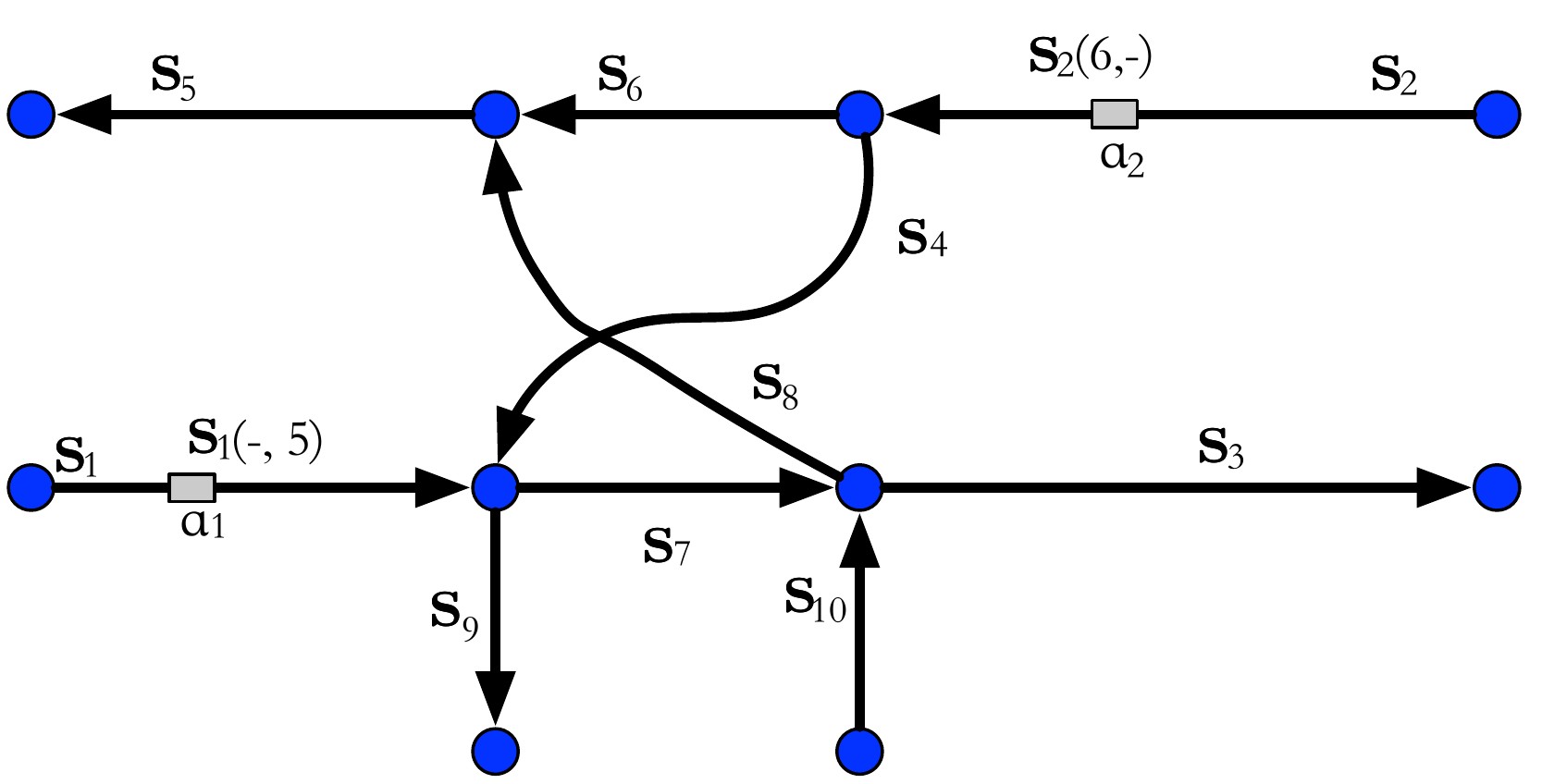} 
\caption{\label{fig:segment} Agents' positions and itineraries}
\end{figure}

For example in Fig.~\ref{fig:segment}, the position of vehicle $a_1$ on the metric graph is $\mathbf{s}_1(-,5)$. 
The segment $\mathbf{s}_1[-,5]$ is the fragment of $\mathbf{s}_1$ from its beginning to the position of $a_1$.


 A ride from a position $p$ to a position $p'$ is a sequence of segments leading from $p$ to $p'$ denoted by $p \stackrel{\mathbf{s}}\rightsquigarrow p'$.
 The distance between two positions $p,~p'$ in a metric graph $G$ is denoted by $d_G$, 
 where 
 \[
 d_G(p,p')=\left\{
 \begin{array}{ll}
      0& \textrm{if~} p=p' \\
      +\infty& \textrm{if~}\not\exists s.p \stackrel{\mathbf{s}}\rightsquigarrow p'\\
      min\{\|\mathbf{s}\|\,|\exists \mathbf{s}.p \stackrel{\mathbf{s}}\rightsquigarrow  p'\} & \textrm{otherwise}
 \end{array}\right.
 \]


\subsubsection{Map to metric graph transformation.}
\begin{figure}[t]
		\centering
		\subfigure[Map\label{fig:map}]{
			\begin{minipage}[t]{0.5\linewidth}
				\centering
				\includegraphics[width=\linewidth]{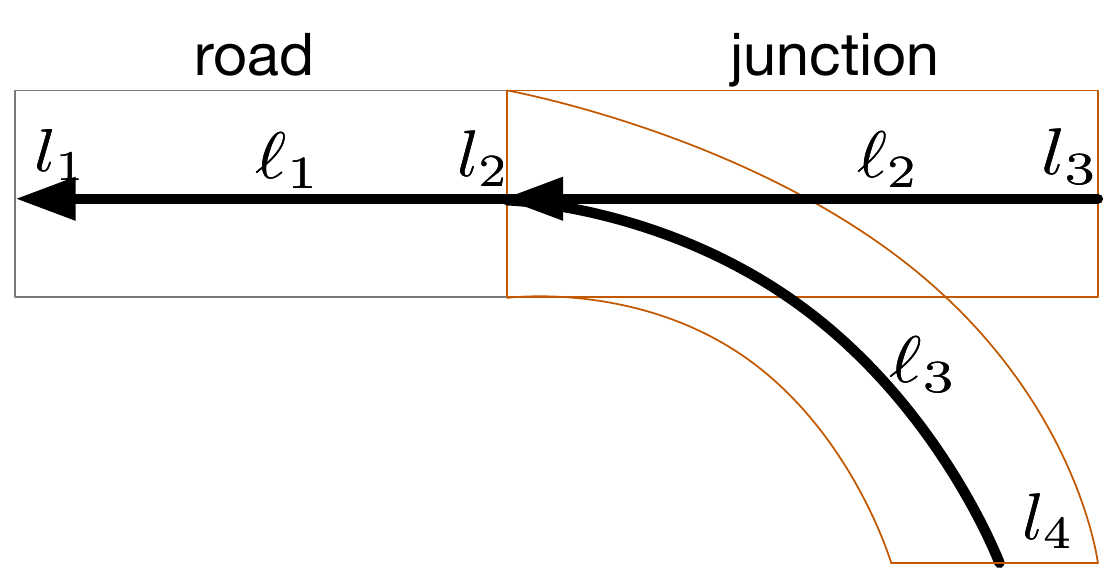}
			\end{minipage}%
		}%
		\subfigure[Metric graph\label{fig:metricgraph}]{
			\begin{minipage}[t]{0.5\linewidth}
				\centering
				\includegraphics[width=\linewidth]{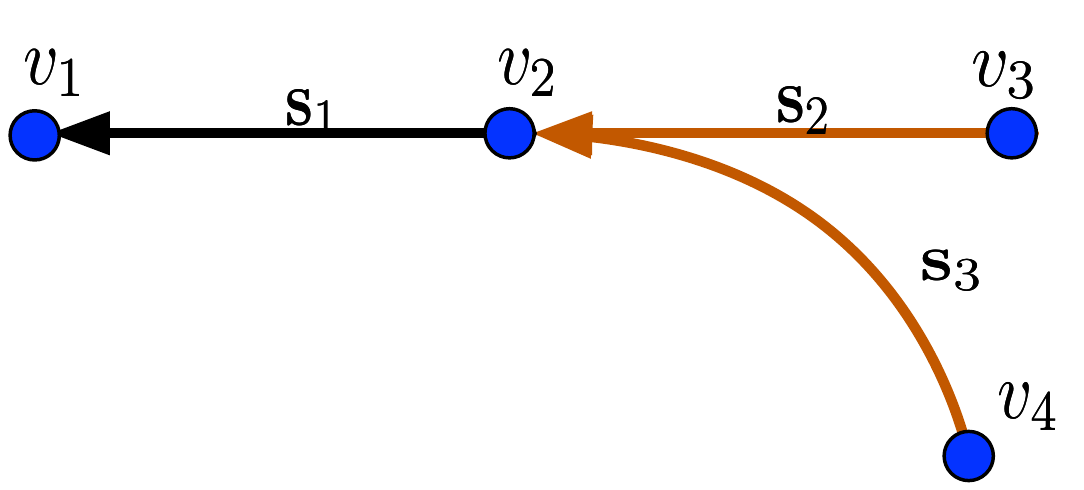}
			\end{minipage}%
		}
		\subfigure[A state of the semantic model\label{fig:scenario-map-example}]{
 			\begin{minipage}[t]{\linewidth}
 				\centering
 				\includegraphics[width=0.95\textwidth]{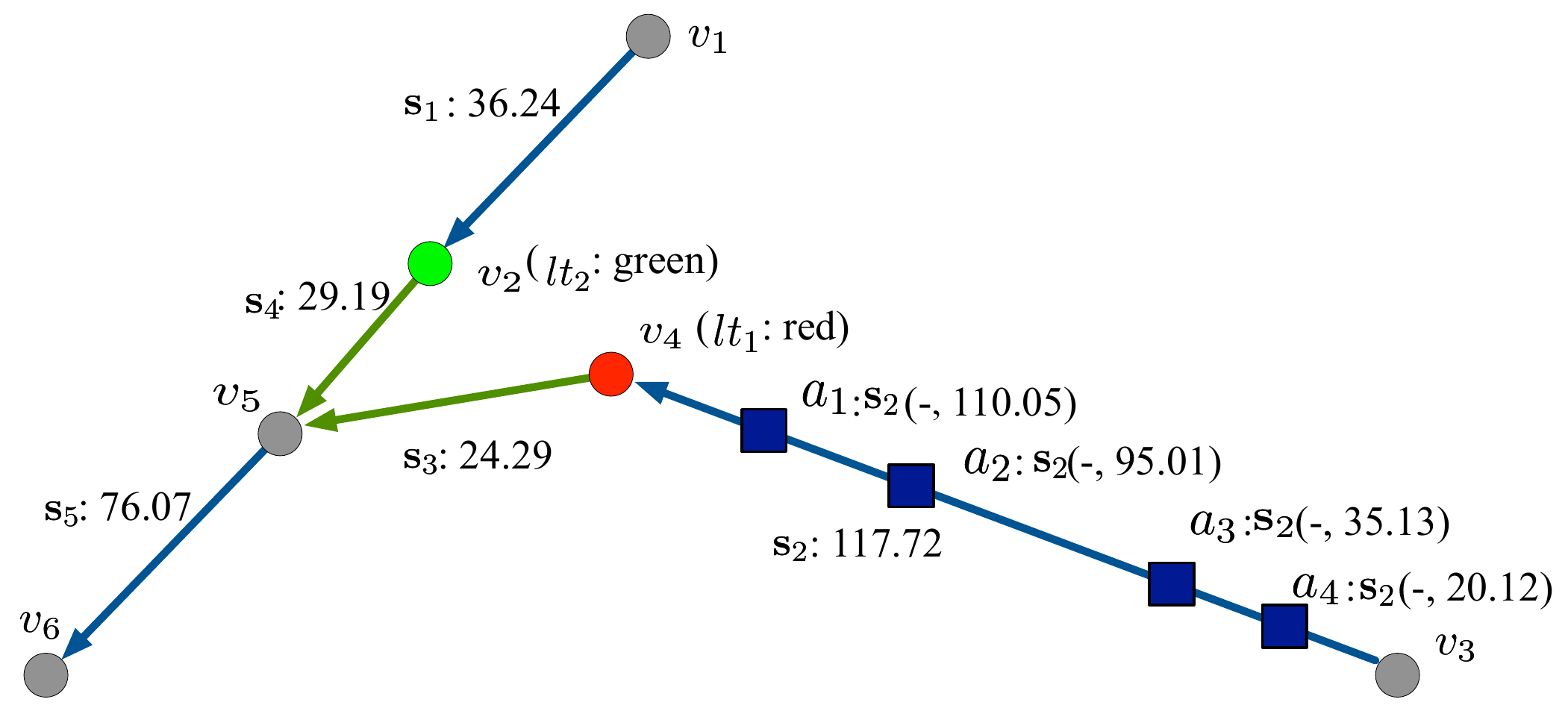}
 			\end{minipage}%
 		}
		\caption{An example map (a), 
  its corresponding metric graph and (b) 
   a state of the semantic model (c)
  \label{fig:mapmodel}}
\end{figure}

The transformation from a map of the Simulator to a metric graph is straight-forward. 
 Each location of a lane is regarded as a vertex, 
 and an edge $e$ with its labelling segment $\mathbf{s}$ is created if there is a lane $\ell$ between two locations.
 Finally, we associate the attributes of the lanes to the corresponding edges and segments, 
 e.g., the turn type of a lane is associated to that of the segment. 
 The traffic signals and stop signs are indicated at their positions on the metric graph.
 For example, the four locations in the map model of Fig.~\ref{fig:map} correspond to the four vertices of the  metric graph in Fig.~\ref{fig:metricgraph}. 
 As there are three connections from $l_2$ to $l_1$, 
 from $l_3$ to $l_2$, 
 and from $l_4$ to $l_2$ respectively, 
 we create three edges labeled with segments $\mathbf{s}_1$, $\mathbf{s}_2$ and $\mathbf{s}_3$ 
 in the corresponding metric graph.

\subsection{Semantic model of an ADS}

In order to simplify the presentation, 
 we regard the mobile traffic participants, 
 such as vehicles and pedestrians as \textit{agents}, 
 and non-mobile traffic participants such as traffic lights and traffic signs as \textit{objects} in the subsequent sections.

The semantic model provides a global view of the dynamics of the simulated system 
 by combining in a coherent way two complementary aspects: 
 on one hand the state of the agents and 
 on the other hand the static environment where they move. 
 We construct the semantic model from the state information of the simulated system exported by the Simulator API.

First, we extract, through the Simulator API, the state information of the agents and objects and build the state of the semantic model.
The state of an ADS consists of a map with the states of its agents and objects. 
The state of an agent includes its position, orientation, velocity and itinerary, 
where an itinerary is a sequence of consecutive segments in the metric graph that cannot begin or finish in a junction. 
 For example, an itinerary for the agent $a_1$ in Fig.~\ref{fig:segment} can be
 $\{ \mathbf{s}_1[8,-] ~ \mathbf{s}_7 ~ \mathbf{s}_3[-$, $7]\}$, and $\mathbf{s}_1[8,-]$ is the first segment of the itinerary.
The state of an object includes its position and for a traffic light additionally its color. 


We represent the state of an agent by $s_a=(it,pos,sp,wt)$ 
 where  $it$ is an itinerary with $it^0$ being the first segment in the itinerary, 
 $pos=it^0(-,0)$ is the position 
 of agent $a$ on the metric graph, 
$sp$ is its speed, 
and $wt$ is the waiting time elapsed since its speed became zero.
 Similarly, the state of a traffic light $lt$ is denoted by $s_{lt}=(pos, color)$, 
 where $pos$ is its position in the metric graph, $color$ can be either red, yellow or green.
 For a stop sign, its state only records its position in the metric graph.
Notice that the generation of the corresponding states of the semantic model requires the computation of  the relative positions of  agents and objects on the metric graph from their physical 2D Cartesian coordinates.

A run of an ADS is a sequence of states generated periodically by the Simulator.
 We assume that the states of agents and objects are updated periodically by the simulation process every time $\Delta t$.
 The global state (state for short) at time $t$ for a set of $n$ agents and objects is denoted by $S_t=(s_1,\ldots,s_n)$.
 The sequence of states in a time interval $[0,m*\Delta t]$ is then denoted by $(S_0,\ldots,S_m)$.
Considering again the map model in Fig.~\ref{fig:mapmodel},
 a state of a simulated system on the corresponding metric graph is shown in Fig.~\ref{fig:scenario-map-example}. 
 There are four vehicles in segment $\mathbf{s}_2$ and two traffic lights in segment $\mathbf{s}_3$ and $\mathbf{s}_4$ respectively. 
 The length of $\mathbf{s}_2$ is 117.72.

The semantic model of a simulated system is then formally defined by the set of all the tuples 
 $(G,\mathbb{S})$, where $G$ is the metric graph of the involved map, 
 and $\mathbb{S}=( {S}_0,\ldots,  {S}_{m})$  
 is a system run extracted from the Simulator in time interval $[0, m*\Delta t]$.
 Each element of a run describes a global state
  including the states of all agents and objects.
 



\section{Property specification and validation}

In this section, we formalize a set of global system properties describing traffic rules as formulas of the linear temporal logic proposed in ~\cite{sifakis-framework} and provide a runtime verification  method to validate such formulas against the semantic model.

\subsection{Property specification}

The metric graph $G=(V,\mathcal{S},E)$ representing a map can be decomposed into two sub-graphs corresponding respectively to its roads and its junctions.

The sub-graph $R=(V_r, \mathcal{S}_r,E_r)$ represents the roads with $E_r\subseteq E$ and $V_r\subseteq V$.
Each road  is a maximal directed path $r=v_0\stackrel{\mathbf{s}_1}\rightarrow v_1\cdots \stackrel{\mathbf{s}_n}\rightarrow v_n$,
where all the vertices $v_1,\ldots,v_{n-1}$ have in-degree and out-degree equal to one. 
Vertices $v_0$ and $v_n$ are called the \textit{entrance} and \textit{exit} of the road $r$, respectively.

The sub-graph $J=(V_\jmath,\mathcal{S}_\jmath, E_\jmath)$ representing the set of junctions is obtained from $G$ by removing from its roads all the vertices except their entrances and exits, with $E_\jmath=E\setminus E_r$  the set of edges labelled with junctions,  $V_\jmath\subseteq V$ the vertices of edges in $E_\jmath$, and $\mathcal{S}_\jmath $ the set of the corresponding segments. 
We assume that in a junction each entrance $v$ has a neighboring exit $v'$ denoted by $v~\texttt{entex}~v'$.



Let $\mathcal{A}$ and $\mathcal{O}$ denote the set of agents and the set of objects, respectively.
The properties of the system are expressed as formulas of a first-order linear temporal logic involving the following types of variables.
 Let $a, o, r, \jmath$ denote the variables of agents, objects, roads, and junctions, respectively. 
 We denote by $x.pos$  
the position of an agent $(x=a)$ or an object $(x=o)$  on the  map, and by $y.en, y.ex$, an entrance and an exit of a road $(y=r)$ or junction $(y=\jmath)$.
Variables $v$ and $e$ represent a vertex and an edge of the map, respectively.
Finally, we introduce three types of predicates on these variables defined as follows:

\begin{itemize}
    \item $v_1 ~~orientation ~~ v_2$, where $orientation \in\{ \texttt{right-of}$, $\texttt{opposite}\}$ describes the relative position between two vertices $v_1$ and $v_2$ in a junction of a map. In particular, predicate $v_1~\texttt{right-of}~v_2$ means that $v_1$ is to the right of $v_2$, and predicate $v_1~\texttt{opposite}~ v_2$ means that $v_1$ and $v_2$ are the origins of two segments oriented in opposite directions. 
    \item $x@y$ indicates that the position of $x$ is at $y$, 
    where $x$ is an agent or object variable and $y$ is a sub-graph restricted to a set of vertices.
    \item $turn(a, d)$ indicates the direction taken by an agent $a$ following its itinerary, 
    where $ d \in \{\texttt{left},\texttt{right}, \texttt{straight}\}$ means respectively that $a$  turns left, right or goes straight.
\end{itemize}

\begin{table*}[]
    \centering
    \begin{tabular}{lcl} 
    \hline
        $(G, \mathbb{S})\models v_1 \texttt{ opposite } v_2$ & iff &
         $v_1 \xrightarrow{\mathbf{s_1}}{v_3} \wedge v_2\xrightarrow{\mathbf{s_2}}{v_4}\wedge (v_1~\texttt{entex}~v_4) \wedge (v_2~\texttt{entex}~v_3)  \wedge \mathbf{s_1}.\tau = \texttt{straight} \wedge \mathbf{s_2}.\tau = \texttt{straight}$ 
        \\
        & & for two vertices $v_3$ and $v_4$ in $G$
        \\
        $(G, \mathbb{S})\models v_1~\texttt{right-of}~v_2$ &
        iff &
        $v_1\xrightarrow{\mathbf{s_1}}{v_3}\wedge v_2\xrightarrow{\mathbf{s_2}}{v_3} \wedge ((\mathbf{s_1}.\tau = \texttt{right} \wedge \mathbf{s_2}.\tau = \texttt{straight})\vee (\mathbf{s_1}.\tau = \texttt{straight} \wedge \mathbf{s_2}.\tau = \texttt{left}))$ \\
        & & for a vertex $v_3$ in $G$
        \\
        $(G, \mathbb{S})\models x@y$ & iff &
        $v_1\xrightarrow{\mathbf{s}}v_2 \wedge s_x.pos = \mathbf{s}(-,\eta)$ for two vertices $v_1, v_2$ and a segment $\mathbf{s}$ in $G$, and a $\eta \in [0, \|\mathbf{s}\|]$, \\
        & & where $s_x$ is the state of agent or object $x$ in $S_0$\\
        $(G, \mathbb{S}) \models turn(a, d)$ & iff & $s_a.it^0.\tau = d$ where $s_a$ is the state of agent $a$ in $S_0$\\
    \hline
    \end{tabular}
    \caption{Semantics of the predicates}
    \label{tab:semantics}
\end{table*}

The semantics of the predicates for elements $(G, \mathbb{S})$ of the  semantic model is defined in Table \ref{tab:semantics}.
 We consider the first order linear temporal logic generated from the above predicates using the temporal modalities $\square, \Diamond, \mathbf{N}, \mathbf{U}$ with their usual meanings.
 For instance, the following formula specifies 
 that for any run of the system with $m$ agents and $k$ objects, 
 formula $\varphi$ should always hold on a system run:
 \[\forall a_1 \cdots \forall a_m .\forall o_1\cdot \forall o_k. \square \varphi(a_1,\ldots, a_m, o_1,\ldots, o_k). \]

To ease the presentation, we also introduce the following predicate that states that the agent $a$ is at an entrance of junction {\rjnote $\jmath$} and heading in direction $d$:
\[take(a, \jmath.en, d) \defeq [a @ \jmath.en \wedge [a @ \jmath.en ~ \mathbf{U} ~ [a @ \jmath\wedge turn(a, d)]]]\]

We provide traffic rules enforcing priorities of agents approaching junctions and their formal specifications in Table~\ref{tab:rules}. 
The first three rules are for junctions with stop signs and the other three are for junctions with traffic lights.
In the rest of this work, we study how to formally validate that these traffic rules are satisfied by the simulated ADS.



\begin{table*}[]
    \centering
    \begin{tabular}{p{17cm}} \hline
     Properties for a stop junction $\jmath$: \\ \hline
     $p_1$: If a vehicle is in the junction, then no other vehicle can be in the junction: \\
     $\forall a. \forall a'.~\square[[a@\jmath \wedge a \neq a']\to \lnot a' @\jmath]$\\
     
    $p_2$: If a vehicle arrives at the same time as another vehicle, the vehicle on the right has the right-of-way:\\
    $\forall \jmath.en.\forall \jmath.en'.\forall a. \forall a'.~\square[[a@\jmath.en \wedge a'@\jmath.en' \wedge a.wt = a.wt' \wedge \jmath.en~\texttt{right-of}~\jmath.en'] \to [[\mathbf{N}~a'@\jmath.en']~\mathbf{U}~a@\jmath]]$\\
    
    $p_3$: The vehicle that arrives first at the entrance will pass  before other vehicles:\\
    $\forall \jmath.en.\forall \jmath.en'.\forall a. \forall a'. \square [[a@\jmath.en \wedge a'@\jmath.en' \wedge a.wt < a'.wt] \to [[\mathbf{N}~a@\jmath.en]~\mathbf{U}~a'@\jmath]]$  \\
    \hline\hline
    Properties for a traffic light junction $\jmath$: \\\hline
     $p_4$: Any vehicle facing a red light must stop until the traffic light turns green, unless the vehicle is turning right. \\
     $ \forall \jmath.en.\forall a. \forall lt.~ \square [[ a@\jmath.en \wedge lt@\jmath.en \wedge lt.cl=red \wedge \lnot take(a, \jmath.en, \texttt{right})] \to 
     [a@\jmath.en ~ \mathbf{U}~lt.cl=green]] $ \\
     
     $p_5$: If a vehicle facing a red light is turning right, then
     the vehicle should wait until there is no vehicle on the left.
     \\
     $ \forall \jmath.en. \forall \jmath.en'. \forall a.\forall a'.\forall lt.~\square[[ a@\jmath.en \wedge a'@\jmath.en' \wedge (\jmath.en ~ 
     \texttt{right-of}~\jmath.en') \wedge lt@\jmath.en \wedge (lt.cl = red) \wedge take(a, \jmath.en, \texttt{right})] \to [[\mathbf{N}~a@\jmath.en] ~\mathbf{U} ~ a'@\jmath]]$ \\
     
     $p_6$: If two vehicles arrive at the entrances of a junction opposite each other and the traffic lights are green, the vehicle turning left must give way to the other.\\
     $ \forall \jmath.en. \forall \jmath.en'. \forall a. \forall a'. \square [[a@\jmath.en \wedge a'@\jmath.en' \wedge \jmath.en~\texttt{opposite}~\jmath.en'\wedge  take(a,\jmath.en,\texttt{left})\wedge \lnot take(a',\jmath.en',\texttt{left})] \to [[\mathbf{N}~a@\jmath.en]~\mathbf{U}~a'@\jmath]]$ \\
    \hline
    \end{tabular}
    \caption{Traffic rules and their formal specifications in linear temporal logic}
    \label{tab:rules}
\end{table*}

\subsection{Runtime verification}

Runtime verification is a lightweight formal verification technique allowing to check whether a system behavior satisfies a desired property based on the observed system runs. 
  Given $(G, \mathbb{S})$ an element of the semantic model constructed from the exported states and a formula $\varphi$ specifying the desired property, 
  we apply  runtime verification as shown in Algorithm~\ref{alg:rv} to check whether $(G, \mathbb{S})$ satisfies the given property. 

\begin{algorithm}
\caption{Runtime verification algorithm}\label{alg:rv}
\begin{algorithmic}[1]
\STATE{Input: a formula $\varphi$ and  an element $(G,\mathbb{S})$ of the  semantic model}
\STATE{Output: verdict showing the verification result}
\STATE{$\varphi'=\textrm{unfold}(\varphi, G, \mathbb{S})$}
\STATE{ $\mathbb{S}^s =\textrm{extract}(\mathbb{S})$}
\STATE{$\varphi''=\textrm{simplify}(\varphi',G, \mathbb{S}^s)$} 
\STATE{$\textrm{verdict}=\textrm{check}(\varphi'', G, \mathbb{S})$}
\end{algorithmic}
\end{algorithm}




The first step applies the \texttt{unfold} function (Line 3), 
 which eliminates all the quantifiers of the input property $\varphi$ and obtains a quantifier-free formula $\varphi'$
 in the following manner.
%
%

Let $D_\kappa$ be the finite domain of a variable $\kappa$.
 Recall that each variable in the formula refers to an element of specific type which can be an agent, object or junction entrance.
 The domain $D_\kappa$ of variable $\kappa$ is a finite set of constants that can be extracted from $(G,\mathbb{S})$ according to the type of $\kappa$.
 For instance, if variable $\kappa$ refers to elements of type \texttt{agent}, 
 its domain $D_\kappa$ will be simply all the agents of the simulated system.

 Quantifier elimination for variable $\kappa$ boils down to the application of the following two rules that instantiate $\kappa$ with all the corresponding elements in its domain $D_\kappa$,
 where $\phi[\kappa/\mathbf{c}]$ represents substitution of $\kappa$  by $\mathbf{c}$ in the formula $\phi$.
 
  
 \begin{align*}
 \forall \kappa.~ \varphi \Leftrightarrow \bigwedge_{\mathbf{c}\in D_{\kappa}}\varphi[\kappa/\mathbf{c}]\\
 \exists \kappa.~ \varphi \Leftrightarrow \bigvee_{\mathbf{c}\in D_{\kappa}}\varphi[\kappa/\mathbf{c}]
 \end{align*}

The function \texttt{extract} (Line 4) constructs a static model $\mathbb{S}^s $,
which contains the state attributes of non-mobile objects 
that do not change over the simulation, e.g., the positions of traffic lights. 
This static model can be used to further simplify the formula.

The function \texttt{simplify} (Line 5) evaluates the predicates involving non-mobile objects in the quantifier-free formula $\varphi'$ according to $(G, \mathbb{S}^s)$, 
and outputs a simplified formula $\varphi''$.

Finally, the function \texttt{check} (Line 6) evaluates the simplified quantifier-free formula $\varphi''$ against $(G,\mathbb{S})$,
 and returns a verdict indicating the satisfaction or not of the given formula.

 For formula evaluation, we adopt the method proposed in \cite{ltl3}, that consists in extracting a deterministic finite state machine characterizing the beavhior specified by formula $\varphi''$.
 Then the satisfiability check boils down to checking the run $\mathbb{S}$ of $(G, \mathbb{S})$ is accepted by the finite state machine.

\section{Rigorous validation using scenarios}

\subsection{Structural equivalence for scenarios}

Given a set of properties $P$, we define an equivalence relation $\sim_P $ on system runs. 
For two system runs $\mathbb{S}_1$, $\mathbb{S}_2$, we put $\mathbb{S}_1 \sim_P \mathbb{S}_2$ when $\mathbb{S}_1 \models p \textrm{ iff } \mathbb{S}_2 \models p$ for each property $p$ of $P$. That is, the equivalence relation $\sim_P $ does not distinguish between system runs that satisfy exactly the same properties. 

To define the structural equivalence for scenarios, we introduce some necessary notations.
Recall that a system state can be defined as a tuple 
$S=(\{s_a\}_{a\in \mathcal{A}},\{s_o\}_{o\in \mathcal{O}})$ 
consisting of states of the agents and objects. 
For a given system state $S$, an abstract scenario is the set of the itineraries of the involved agents of $\mathcal{A}$, and denoted by $as(S)=\{it_a\}_{a\in \mathcal{A}}$. For example, the abstract scenario for the set of agents $\{a_1, a_2\}$ in Fig.~\ref{fig:segment} is 
$\{a_1: \mathbf{s}_1[8,-] \mathbf{s}_7 \mathbf{s}_3[-$, $7]$, $a_2: \mathbf{s}_2[6,-]  \mathbf{s}_4 \mathbf{s}_9[-,6]\}$.
%
%

Clearly, any state $S$ can be considered as the disjoint union $S=as(S)\oplus cn(S)$ where $cn(S)$ is the context of S including the dynamic attributes of the states. This decomposition is motivated by the need to distinguish between the part of the agent's state that remains unchanged and the part that changes during the evolution of the system. 

Now, we define a structural equivalence $\approx_P$ on abstract scenarios parameterized by the set of properties $P$. For two abstract scenarios  $as_1, as_2$, we say that $as_1 \approx_P as_2$ if $\mathbb{S}_1 \sim_P \mathbb{S}_2$ for any runs $\mathbb{S}_1, \mathbb{S}_2$  of the simulated system from the initial states $S_1$ and $S_2$ such that  $as(S_1) =as_1, as(S_2)= as_2$ and $cn(S_1)=cn(S_2)$. 
That is, two equivalent abstract scenarios extended with identical contexts define two initial states from which the simulated system will generate equivalent runs with respect to the properties of $P$.

In our testing methodology, we assume that the scenario generator extends abstract scenarios by providing each agent its initial state and determining the whole initial execution context.

We focus on abstract scenarios of a system with a map and a fixed set of agents and objects. A key observation is that for most traffic rules, their application depends mainly on the relative positions of the agent itineraries regardless of the agent dynamics. For instance, if the itineraries of two vehicles cross a junction, the  traffic rules in Table~\ref{tab:rules} are applicable independently of their speeds.

The considered abstract scenarios may involve multiple roads and junctions. As the behavior of a vehicle in one junction has little influence on its behavior in subsequent junctions of its itinerary, we focus on testing policies for different types of junctions.

\subsection{Coverage criteria for junctions}

We discuss below principles for generating abstract scenarios for junctions guided by a coverage criterion exactly as for structural testing of software systems~\cite{basili1987comparing}.
 We show how structural equivalence on abstract scenarios can be computed for the considered two types of junctions and the corresponding properties. 
 We assume, without loss of generality, that the same type of traffic rules are applied to all junction entrances, i.e., regulation either by stop signs or by traffic lights.

Considering a junction with $k+1$ entrances and corresponding exits, 
we can encode possible itineraries and the corresponding abstract scenarios for vehicles that are at its entrances in the following manner using a set of $k$ symbolic directions $\{d_1, \ldots,d_k\}$. 
For a vehicle entering at entrance $i$, the symbolic direction $d_j$ means that the vehicle is heading to exit $i+j~(mod~ (k+1))$. 
For instance, for a vehicle at entrance $0$, its possible itineraries can be represented by the ordered set $\{0_{d_1}, 0_{d_2}, \ldots, 0_{d_k}\}$ heading to exits from $1$ to $k$. 
Thus if we consider maximal abstract scenarios with one vehicle per entrance of the junction, we will have $k^{(k+1)}$ different abstract scenarios (for each entrance we have $k$ different itineraries and we have $k+1$ different entrances).
The general form of a maximal abstract scenario is $i_{1 d_{j_1}}\cdots i_{(k+1) d_{j_k}}$, 
where $j_1, j_2,..., j_k$ take values in the set $\{1,2,...,k\}$, 
and all the entrance indexes $i_1, ..., i_{(k+1)}$ are different.

\begin{figure}
    \centering
    \includegraphics[width=0.35\textwidth]{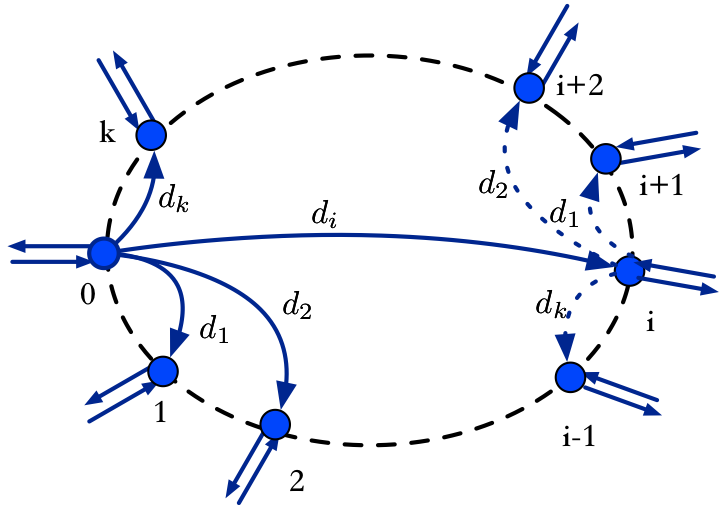}
    \caption{A  junction with $k+1$ entrances and exits }
    \label{fig:kentrance}
\end{figure}

Note that if the traffic rules do not particularize the inputs of the junction according to their positions, 
 one can easily define structurally equivalent abstract scenarios by simply rotating the inputs and preserving the symbolic directions. 
 This is generally the case when access to an junction  is governed by the same traffic rules, such as an all-way stop or a traffic signal controlled junction. 
 %
Thus, the abstract scenario $i_{1 d_{j_1}}\cdots i_{(k+1) d_{j_k}}$ under these assumptions is structurally equivalent to the one $(i_1+1)_{d_{j_1}}\cdots (i_{(k+1)} +1)_{d_{j_k}}$  where the sum operation on indices is modulo $k+1$.  
That is, this transformation gives structurally equivalent abstract scenarios by simple rotation of the positions of the agents 
without changing the relative positions of their trajectories. 

\begin{figure}
    \centering
    \includegraphics[width=0.45\textwidth]{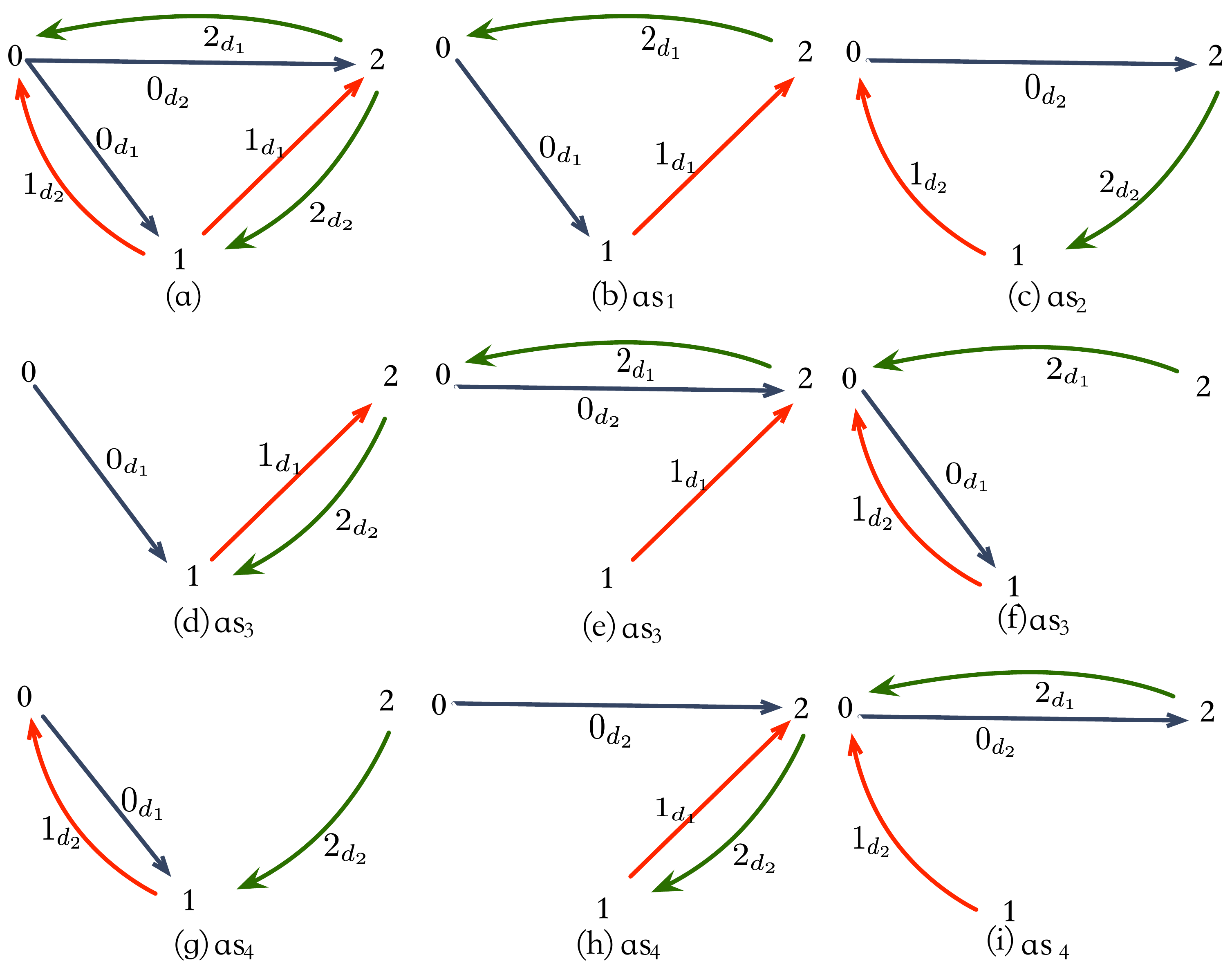}
    \caption{A 3-way junction and its equivalent classes }
    \label{fig:eqclasses}
\end{figure}

In Fig.~\ref{fig:eqclasses} we show how possible itineraries 
and the corresponding abstract scenarios can be generated for a junction with $3$ entrances/exits (Fig.~\ref{fig:eqclasses}.(a)). 
We have $k=2$, thus we have $2^3 = 8$ abstract scenarios that are shown and grouped in four equivalence classes ($as_1$: Fig.~\ref{fig:eqclasses}(b), $as_2$: Fig.~\ref{fig:eqclasses}(c), $as_3$: Fig.~\ref{fig:eqclasses}(d)-(f), and $as_4$: Fig.~\ref{fig:eqclasses}(g)-(i)). The abstract scenario $0_{d_1} 1_{d_1} 2_{d_1}$ is alone in its equivalence class as well as the abstract scenario $0_{d_2} 2_{d_2} 1_{d_2}$. The abstract scenario $0_{d_1} 1_{d_1} 2_{d_2}$ is equivalent to two other abstract scenarios: $1_{d_1} 2_{d_1} 0_{d_2}$ and $2_{d_1} 0_{d_1} 1_{d_2}$ depicted in the second row of the figure. Finally, the third row presents three other structurally equivalent abstract scenarios. 

Similarly for four-way junctions we have $3^4$ different maximal abstract scenarios, some of which are alone in their equivalence class and others belong to classes with a maximum number of four elements. For the sake of space, we omit the elaborations.

\subsection{Generating concrete scenarios}

By following an approach inspired from the metamorphic testing~\cite{chen2018metamorphic}, we use equivalent scenarios to test the simulated system with respect to the properties under consideration. Equivalent scenarios are obtained by extending equivalent abstract scenarios with the same dynamic attributes. 

Note that given two abstract scenarios for a junction, we can extend them into equivalent abstract scenarios by appending segments that modify corresponding itineraries in the same manner. That is, given two abstract scenarios $as_1=(it_{11}, \ldots, it_{1n})$ and $as_2=(it_{21},\ldots, it_{2n})$, such that $as_1 \approx as_2$, we can extend them  by appending road segments $\mathbf{s}_1,\ldots, \mathbf{s}_n$ leading to the entrances of the junction that are the starting points of the itineraries. We thus  obtain abstract scenarios $as_1' =(\mathbf{s}_1it_{11},\ldots, \mathbf{s}_n it_{1n})$ and $as_2' = (\mathbf{s}_1 it_{21}, \ldots, \mathbf{s}_n it_{2n})$ such that $as_1' \approx as_2'$. We use this simple extension mechanism to generate new equivalence classes of abstract scenarios by simply changing a parameter that is the distance of corresponding itineraries from the entrance of the junction.

For each abstract scenario, we consider one vehicle per direction. The vehicles are initialized at different positions and with different speeds in order to generate concrete scenarios. The positions of the vehicles cover different distances from the entrance of a junction. 
The speed of the vehicles ranges from zero to the speed limit of the road on which they are located.
However, in order to generate realistic scenarios we should choose the initial speeds and distances from the entrance of a junction in such a manner that the vehicle can safely stop before the entrance of the junction if needed. According to the vehicle dynamics, for each distance $d$, there is a maximal braking speed $v(d)_{max}$ beyond which the vehicle cannot safely stop by braking over the distance $d$. Thus, we will consider scenarios where the following condition holds: if the distance from the entrance of a junction is $d$, then the initial speed of the vehicle will be less than or equal to $v(d)_{max}$. In that manner, we exclude the cases where a vehicle may violate a traffic rule for a junction because of excessive speed. 
Since $v(d)_{max}$ is an increasing function of $d$, the correspondence between the distance and the maximal braking speed can be estimated by iterative testing of the vehicle dynamics in the Simulator.

\section{Implementation and experimentation}

\subsection{Implementation}

\begin{figure}[h]
\centering
\includegraphics[width=0.49\textwidth]{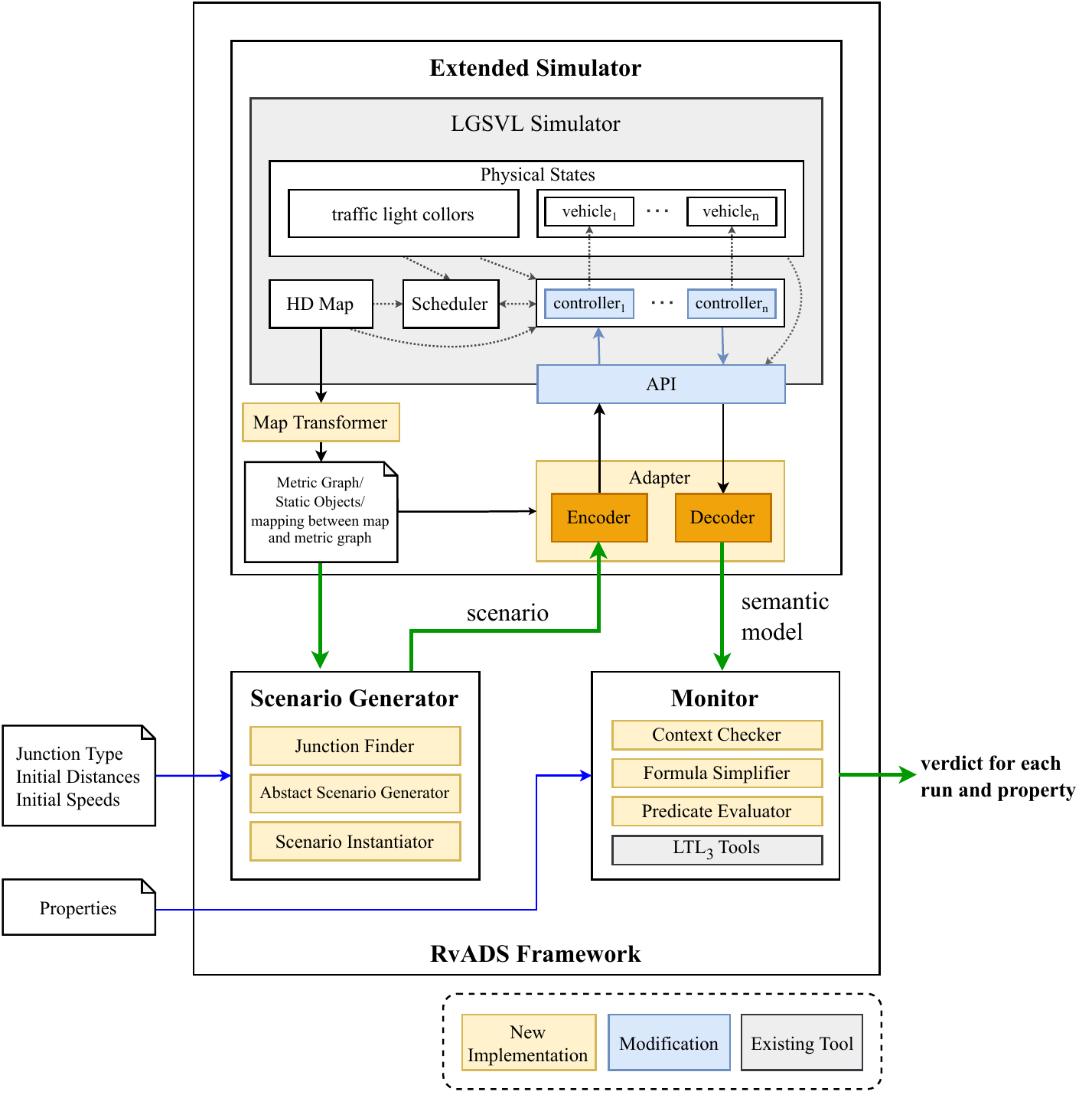}
\caption{\label{fig:svl} RvADS  with the integrated LGSVL Simulator}
\end{figure}

We have implemented in the RvADS framework the proposed rigorous testing method to validate autonomous driving systems.
RvADS shown in Fig.~\ref{fig:svl} integrates a Scenario Generator and a Monitor with an ADS simulator.
The Scenario Generator produces equivalent scenarios obtained from abstract scenarios that cover all the maximal possible configurations of vehicles from a junction entrance.
The Monitor performs  runtime verification of the simulated system against a set of properties specifying traffic rules applied at junctions.

RvADS uses the LGSVL Simulator to simulate an ADS.
 The LGSVL Simulator provides a controller for each vehicle,
 which is in charge of the speed control, as well as the direction to cross the junction.
 For speed control, 
 the LGSVL controller evaluates the target speed of the vehicle according to the following information inside the simulator:
 \begin{itemize}
    \item Physical states:  concerning the traffic light state, vehicle positions and speeds.
    \item HD Map: describing the junctions and their structure, the length of the roads, the relationship between roads and junctions, and the positions of objects (i.e., traffic light and stop sign) on roads.
    \item Scheduling policy:  taking into account the order of arrival of vehicles at each stop sign junction and deciding their order of entry in sequence according to the HD Map and the physical states.
\end{itemize}

In addition to the speed control, 
the controller randomly decides  a direction to follow when a vehicle is approaching a junction.
Furthermore, as explained, the original LGSVL Simulator API provides only the physical coordinates of the vehicles, without any explicit connection to the map information.
In order to make it compatible with the proposed semantic model construction and validation, we have also extended the LGSVL Simulator.
 The major modifications are as follows:

\begin{itemize}
\item First, we have modified the implementation of the controllers
to allow them to follow an  itinerary on the HD Map
when the vehicle approaches a junction. The itinerary 
on the HD Map is defined as a sequence of segments using the extended API.
In addition, we have established a connection between the
controllers and the API such that they can access both the position and the itinerary of a vehicle on the HD map.

\item Second, we have developed a Map Transformer from the HD map and static object state information in the Simulator into the corresponding metric graph and object state representation. 

                
\item Third, we have developed an Adapter between the Extended Simulator, the Scenario Generator, and the Monitor.
Based on the mapping between the HD Map and the
metric graph, the encoder in the Adapter transforms an
itinerary on the metric graph into an itinerary on the
HD Map;  the decoder transforms state information of the simulated system into a state of the semantic model.
\end{itemize}

The Scenario Generator receives inputs specifying junction types, and parameters of abstract scenarios such as initial positions and speeds of the vehicles.  It uses  a Junction Finder to find a specific junction of the metric graph according to the input junction type. Then, it generates scenarios by extending abstract scenarios for the junction with the  attributes provided in the input.


The Monitor first checks whether the semantic model is relevant to the context of the given property or not. 
There are two cases where the model is not relevant: 
 1) the property specifies a rule that is not applicable to the scenario of the given semantic model, 
 e.g., rules for a traffic light are not applicable to a four-way stop;
 2) the property is trivially satisfied, 
 e.g., when a property specifies that if vehicles arrive at a junction at the same time, the vehicle on the right has the right-of-way, 
 it is trivially satisfied if there are no vehicles arriving at the same time.
 In such cases, the verdict "NA" is given directly.

Otherwise, the monitor checks the satisfaction of the property.
It first eliminates the universal quantifiers and transforms the property specification into a propositional linear temporal logic formula using Formula Simplifier, where each atomic predicate of the property is converted into a proposition that is evaluated on the semantic model by the Predicate Evaluator.
Then the LTL$_3$ tool ~\cite{ltl3} is used to convert the propositional linear temporal logic formula into a finite state machine (FSM) to check whether the property is satisfied by the semantic model and output the verdict.

\subsection{Experimental setup}

\begin{figure}[tp]
		\centering
				\includegraphics[width=0.5\linewidth]{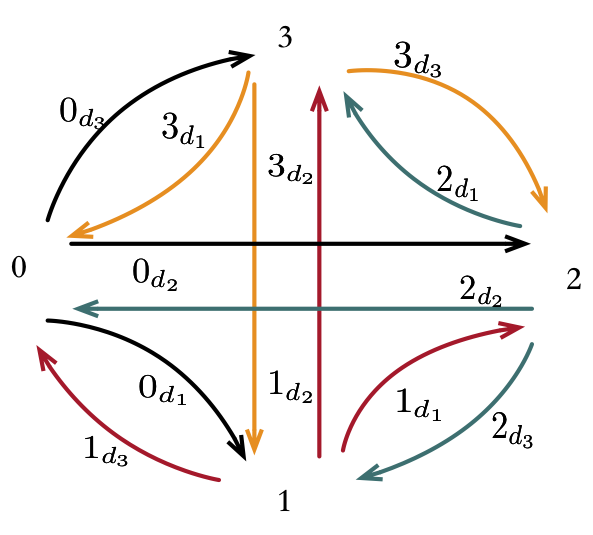}
		\caption{Possible itineraries for a 4-way junction \label{fig:4way}}
\end{figure}

We consider 4-way junctions (as shown in Fig.~\ref{fig:4way}) equipped with stop signs or traffic lights as the static environment of the simulated system. 
For the stop sign junction, we consider that the  stop signs are located exactly at the entrances.
A 4-way junction involves 81 abstract scenarios, which are partitioned into 24 structural equivalence classes.
For each scenario, we set one vehicle per entrance of the junction. 
 
For the simulation of each scenario, 
 the modified LGSVL Simulator can enforce the execution of the given scenario so that each vehicle follows the corresponding itinerary. 
 RvADS checks whether the properties specifying the traffic rules presented in Table~\ref{tab:rules} are satisfied by the simulated system at runtime. 
 If the semantic model of a scenario is not within the context of the property, then RvADS produces an "NA" verdict.
Otherwise if the generated system run satisfies a property, 
RvADS delivers a "pass" verdict; else it delivers a "fail" verdict.

\begin{figure}[tp]
    \centering
     \includegraphics[width=0.98\linewidth]{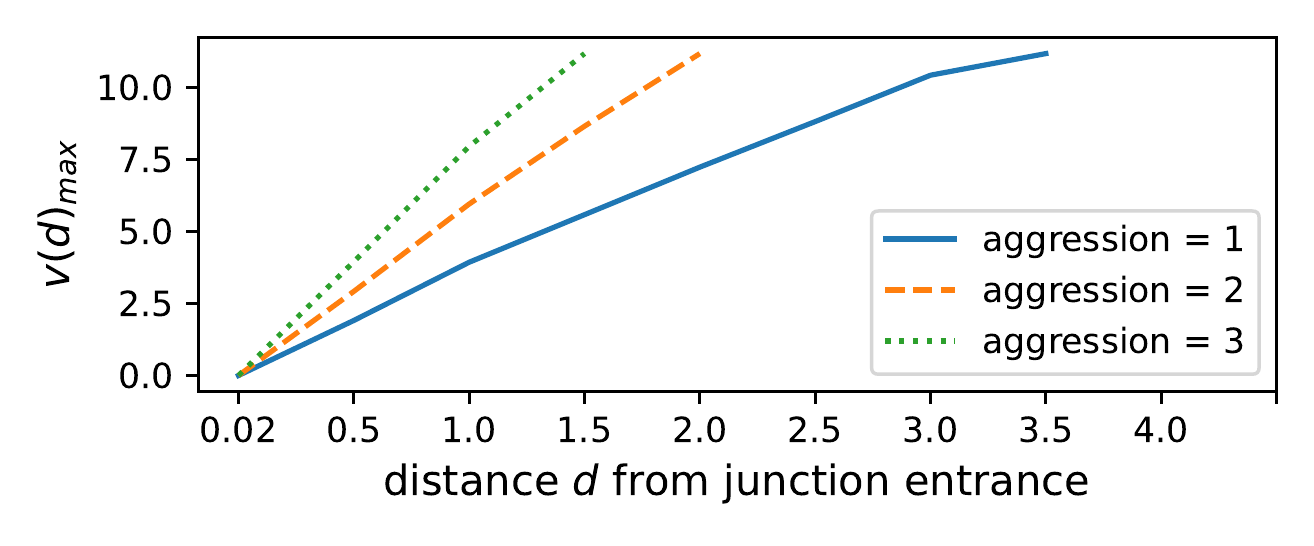}
	\caption{
     Maximal safe speeds as a function of the braking distance 
	\label{fig:safe-speed}}
\end{figure}

Recall that the agents are assigned with various initial speeds and distances from the junction entrances to generate concrete scenarios.
 In order to generate realistic scenarios, 
 the distance from a stop sign and the maximal braking speed should respect certain relations as explained in the previous section.
 In Fig.~\ref{fig:safe-speed},  we plot the maximal safe braking speed (in meters/second) as a function of the distance (in meters) from the entrance of a junction. 
 We consider a city map with a speed limit of 11.176 m/s. 
 The three curves correspond to the three speed adjustment rates 
 (i.e., aggression $\in [1, 2, 3]$) used by the LGSVL Simulator.
 For a given distance,
 the lower the aggression,  the lower the safe braking speed.
 We set aggression = 1 in this experimentation.

We remark that in estimating the correspondence between the maximal braking speed and the distance from a stop sign, we have found that a vehicle cannot brake at the stop sign when the distance is less or equal to 0.01 meters to the stop sign, even if the initial speed is zero. The reason is that the controller of the LGSVL Simulator will always apply a speeding-up policy to initialize the vehicle without checking the safe braking distance to the stop sign. That reveals the lack of controllability for small distances in the Simulator that would occur even when simulating a single vehicle.

\subsection{Experimentation}

RvADS proves to be very effective at testing ADS and allows systematic exploration of high-risk situations whose probability of occurrence is extremely low in random testing. In particular, we have been using RvADS to uncover several deficiencies of the LGSVL Simulator as summarized in Table \ref{table-issues}.

\begin{table*}[htp]
\begin{adjustbox}{width=0.95\textwidth,center}
\setlength\tabcolsep{1pt}
    \centering
    \begin{tabularx}{\textwidth}{c|c|Y@{\hskip 0.2in}Y@{\hskip 0.2in}Y||*{3}{Y}||*{3}{Y}|*{3}{Y}||*{3}{Y}}
         \hline
         \multirow{3}{*}{\#} &
         class of
         & \multicolumn{3}{c||}{distance (0.01,0.01,0.01,0.01)}
         & \multicolumn{3}{c||}{distance (0.3,0.3,0.3,0.3)} &\multicolumn{6}{c||}{distance (20,20,20,20)}
         & \multicolumn{3}{c}{distance (0.3,0.3,20,20)}\\\cline{3-17}
        & abstract &\multicolumn{3}{c||}{A: speed (0,0,0,0)} &\multicolumn{3}{c||}{B: speed (0,0,0,0)}& \multicolumn{3}{c|}{C: speed (0,0,0,0)} &\multicolumn{3}{c||}{D: speed (10,10,10,10)}& \multicolumn{3}{c}{E: speed (0,0,0,0)} \\\cline{3-17}
        & scenarios& $p_1$ & $p_2$ & $p_3$ & $p_1$ & $p_2$ & $p_3$ & $p_1$ & $p_2$ & $p_3$ & $p_1$ & $p_2$ & $p_3$& $p_1$ & $p_2$ & $p_3$ \\\hline

\multirow{4}{*}{1} 
&$0_{d_2}1_{d_1}2_{d_1}3_{d_1}$& Fail & Fail & NA & Fail& Fail& NA& Fail& Fail& NA& Fail& Fail& NA& Fail& Fail& Pass\\
&$0_{d_1}1_{d_2}2_{d_1}3_{d_1}$& Fail & Fail & NA & Fail& Fail& NA& Fail& Fail& NA& Fail& Fail& NA& Fail& Fail& Pass\\
&$0_{d_1}1_{d_1}2_{d_2}3_{d_1}$& Fail & Fail & NA & Fail& Fail& NA& Pass& Fail& NA& Pass& Fail& NA& Pass& Fail& Pass\\
&$0_{d_1}1_{d_1}2_{d_1}3_{d_2}$& Fail & Fail & NA & Fail& Fail& NA& Fail& Fail& NA& Fail& Fail& NA& Fail& Fail& Pass\\
\hline
\multirow{4}{*}{2} 
& $0_{d_3}1_{d_1}2_{d_1}3_{d_1}$& Fail & Fail & NA & Fail& Fail& NA& Fail& Fail& NA& Fail& Fail& NA& Fail& Fail& Pass\\
&$0_{d_1}1_{d_3}2_{d_1}3_{d_1}$& Fail & Fail & NA & Fail& Fail& NA& Fail& Fail& NA& Fail& Fail& NA& Fail& Fail& Pass\\
&$0_{d_1}1_{d_1}2_{d_3}3_{d_1}$& Fail & Fail & NA & Fail& Fail& NA& Pass& Fail& NA& Pass& Fail& NA& Fail& Fail& Pass\\
&$0_{d_1}1_{d_1}2_{d_1}3_{d_3}$& Fail & Fail & NA & Fail& Fail& NA& Fail& Fail& NA& Fail& Fail& NA& Fail& Fail& Pass\\
\hline
    \end{tabularx}
    \end{adjustbox}
    \caption{Experimental results for a 4-way stop junction
    }
    \label{tab:eq4waystop}
\end{table*}

\begin{table*}[htp]
\begin{adjustbox}{width=0.95\textwidth,center}
\setlength\tabcolsep{1pt}
    \centering
    \begin{tabularx}{\textwidth}{c|c|Y@{\hskip 0.2in}Y@{\hskip 0.2in}Y||*{3}{Y}||*{3}{Y}|*{3}{Y}||*{3}{Y}}
         \hline
         \multirow{3}{*}{\#} & class of
         & \multicolumn{3}{c||}{distance  (0.01,0.01,0.01,0.01)} 
          & \multicolumn{3}{c||}{distance  (0.3,0.3,0.3,0.3)} &\multicolumn{6}{c||}{distance  (20,20,20,20)}&\multicolumn{3}{c}{distance  (20,0.3,20,0.3)}
          \\\cline{3-17}
        & abstract 
        &\multicolumn{3}{c||}{F: speed (0,0,0,0)}
        &\multicolumn{3}{c||}{G: speed (0,0,0,0)}
        & \multicolumn{3}{c|}{H: speed (0,0,0,0)} &\multicolumn{3}{c||}{I: speed (10,10,10,10)}
         &\multicolumn{3}{c}{J: speed (0,0,0,0)}\\\cline{3-17}
          
          & scenarios& $p_4$ & $p_5$ & $p_6$ & $p_4$ & $p_5$ & $p_6$ & $p_4$ & $p_5$ & $p_6$ & $p_4$ & $p_5$ & $p_6$ & $p_4$ & $p_5$ & $p_6$ \\\hline
          \multirow{4}{*}{3}
        &$0_{d_1}1_{d_2}2_{d_3}3_{d_3}$&Fail&NA&Fail& Pass& NA& Fail& Pass& NA& Fail& Pass& NA& Fail& Pass& NA& Fail\\  &$0_{d_3}1_{d_1}2_{d_2}3_{d_3}$&Fail&NA&Fail& Pass& NA& Fail& Pass& NA& Fail& Pass& NA& Fail& Pass& NA& Fail\\
&$0_{d_3}1_{d_3}2_{d_1}3_{d_2}$&Fail&NA&Fail& Pass& NA& Fail& Pass& NA& Fail& Pass& NA& Fail& Pass& NA& Fail\\
&$0_{d_2}1_{d_3}2_{d_3}3_{d_1}$&Fail&NA&Fail& Pass& NA& Fail& Pass& NA& Fail& Pass& NA& Fail& Pass& NA& Fail\\
\hline
\multirow{4}{*}{4 }  & $0_{d_2}1_{d_1}2_{d_2}3_{d_1}$&NA&Pass&NA& NA& Pass& NA& NA& NA& NA& NA& NA& NA& NA& Fail& NA\\
&$0_{d_1}1_{d_2}2_{d_1}3_{d_2}$&NA&Pass&NA& NA& Pass& NA& NA& NA& NA& NA& NA& NA& NA& Fail& NA\\
&$0_{d_2}1_{d_1}2_{d_2}3_{d_1}$&NA&Pass&NA& NA& Pass& NA& NA& NA& NA& NA& NA& NA& NA& Fail& NA\\
& $0_{d_1}1_{d_2}2_{d_1}3_{d_2}$&NA&Pass&NA& NA& Pass& NA& NA& NA& NA& NA& NA& NA& NA& Fail& NA\\
\hline
    \end{tabularx}
    \end{adjustbox}
    \caption{Experimental results for a 4-way traffic light junction
    }
    \label{tab:lt-eqnew}
\end{table*}

\begin{table*}
\SetTblrInner{rowsep=0pt}
\begin{tblr}{
  colspec = {|c|X[5,j]|}, rowspec={Q[m]Q[m]Q[m]Q[m]Q[m]Q[m]}
}
    \hline
    Deficiency ID & Explanation \\
    \hline
    {\bf I$_1$} & Lack of controllability for short distances. During initialization, the controller always randomly assigns a rate for vehicle acceleration without taking into account the safe braking distance ahead.
When the distance to go is small, the vehicle may not be able to brake safely. This may happen even for a single vehicle.  
    \\
    \hline
    {\bf I$_2$} & Hidden guidance for control. Vehicle control uses a boundary zone for junctions that cannot be obtained by sensing the junction environment delineated by entrances/exits and traffic signs. 
    \\
    \hline
    {\bf I$_3$} & No consideration of priorities between different itineraries with the same waiting time. The Scheduler sets priorities according to the creation order of vehicles.
    \\ 
    \hline
    {\bf I$_4$} & No consideration of the priorities when turning right at a red light. When a right turning vehicle faces a red light and stops at the entrance of a junction, it does not consider the priority of the vehicle at its left side facing a green light.\\ 
    \hline
    {\bf I$_5$} & Application of a partial order between the lanes of a junction followed by the vehicles. However, this order is incomplete  and leaves unresolved conflicts. For example, for two vehicles coming from opposite directions, the vehicle turning left can enter the junction before the other one turning right.  \\ 
    \hline
\end{tblr}
\caption{Deficiencies discovered in the LGSVL Simulator}
\label{table-issues}
\end{table*}

\subsubsection{Experimentation with a 4-way stop junction.}

A 4-way junction contains 24 structural equivalence classes. 
%
%
In Table~\ref{tab:eq4waystop} we provide testing results for the traffic rules $p_1, p_2$, and $p_3$ for 8 abstract scenarios grouped into two structural equivalence classes.
The concrete scenarios are then obtained by setting different distances from the entrance (i.e., 0.01m, 0.3m and 20m) and different initial speeds (i.e., 0m/s and 10 m/s) that respect the safe braking distance relation as shown in Fig.~\ref{fig:safe-speed}.
For brevity, we denote a scenario by the index of its abstract scenario and the index of the initial distance and speed configuration. For example, $(1,A)$ includes the scenarios in the first structural equivalence class with 0.01 meters from the stop sign and zero initial speed for each vehicle. 
We detail below the causes for each property violation.
Experiments show that, for an all-way stop junction, vehicles do not decide autonomously based on knowledge of their surroundings. The Simulator uses a Scheduler to decide which vehicle should move first in case of a conflict. In addition, the Scheduler considers the junction as a different zone than the one defined by its entrances and exits. The use of such a centralized control mechanism goes against the assumption that vehicles are autonomous. Furthermore as discussed below, it is a source of inconsistencies as the Scheduler's perception of the junction  differs from that of a vehicle based on its own sensor equipment.   





\begin{figure}[t]
        \subfigure[Violating property $p_1$  \label{fig:issue1}]{
			\begin{minipage}[t]{0.45\linewidth}
				\centering				\includegraphics[width=0.93\linewidth]{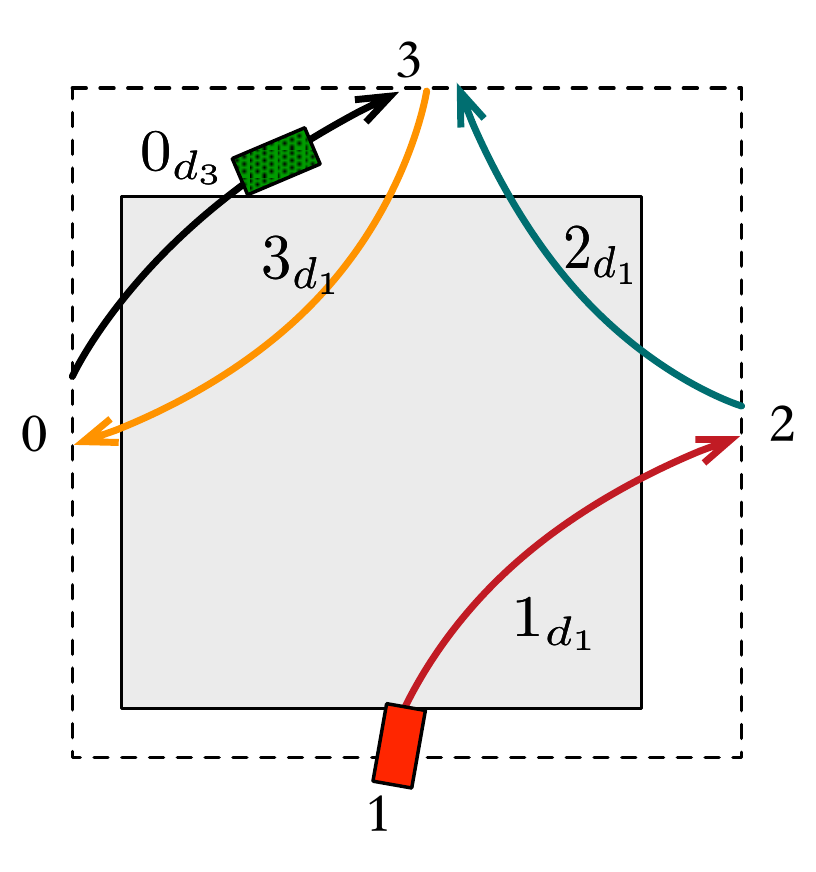}
			\end{minipage}%
		}%
	\subfigure[Satisfying property $p_1$  \label{fig:issue2}]{
			\begin{minipage}[t]{0.45\linewidth}
				\centering				\includegraphics[width=0.98\linewidth]{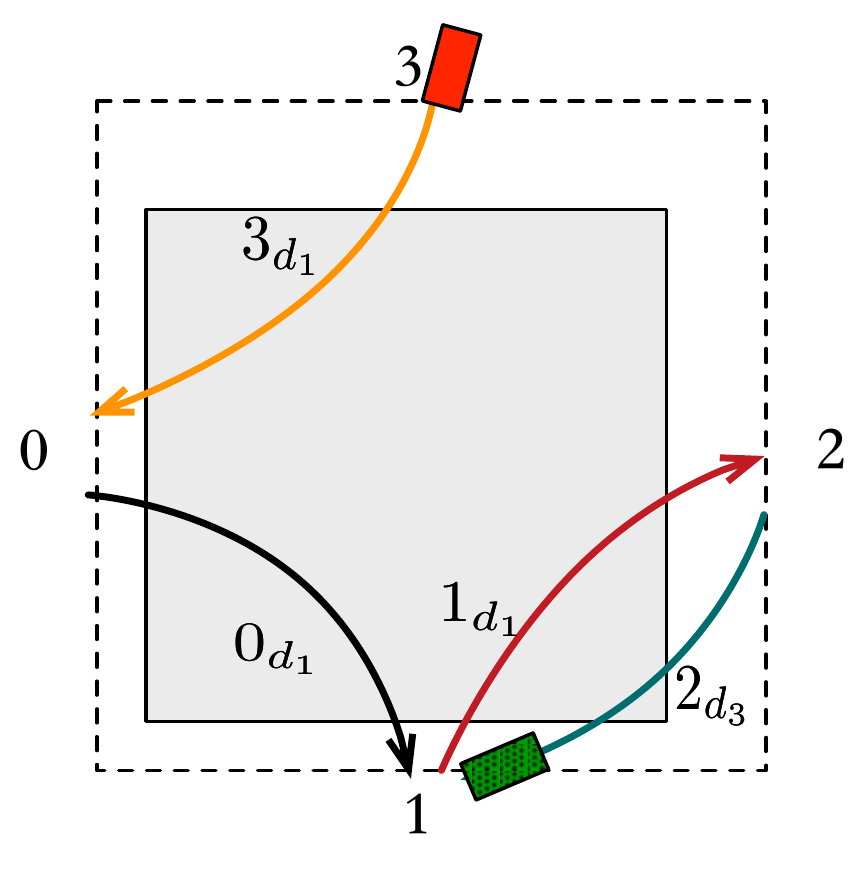}
			\end{minipage}%
		}%
  \\
		\subfigure[Violating property $p_5$  \label{fig:issue3}]{
			\begin{minipage}[t]{0.45\linewidth}
				\centering				\includegraphics[width=0.95\linewidth]{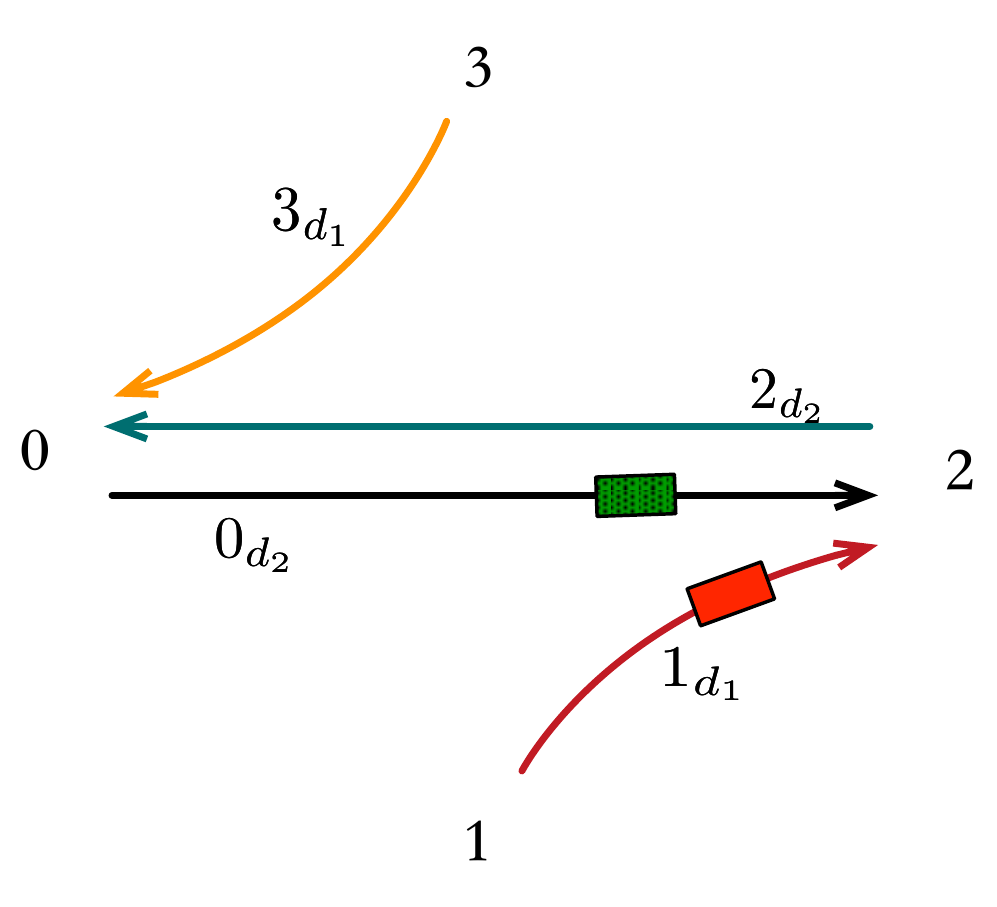}
			\end{minipage}%
		}%
		\subfigure[Violating property $p_6$  \label{fig:issue4}]{
			\begin{minipage}[t]{0.45\linewidth}
				\centering				\includegraphics[width=0.95\linewidth]{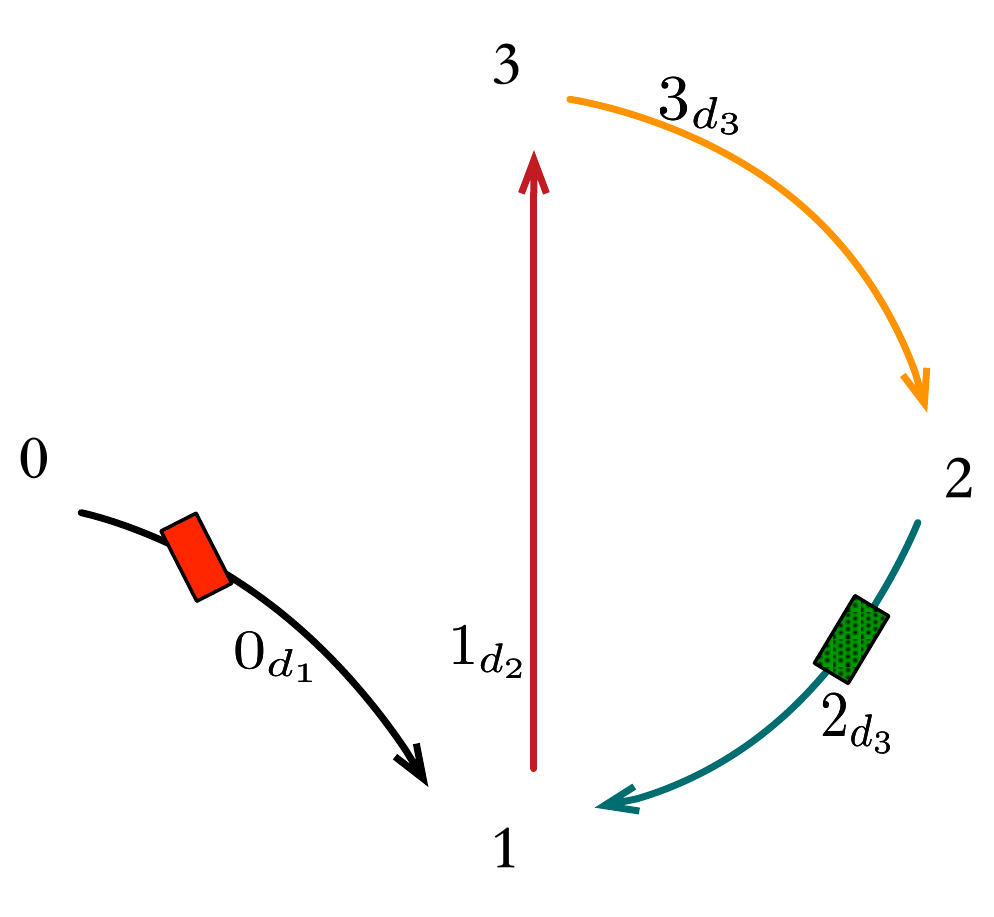}
			\end{minipage}%
		}%
		
\caption{Example scenarios  \label{fig:issues}
	}
\end{figure}

\begin{figure*}[h!]
	\centering
	\subfigure[Violation of $p_2$ ~(at time $t-1$)\label{fig:simp2}]{
			\begin{minipage}[t]{0.45\linewidth}
				\centering				\includegraphics[width=0.91\linewidth]{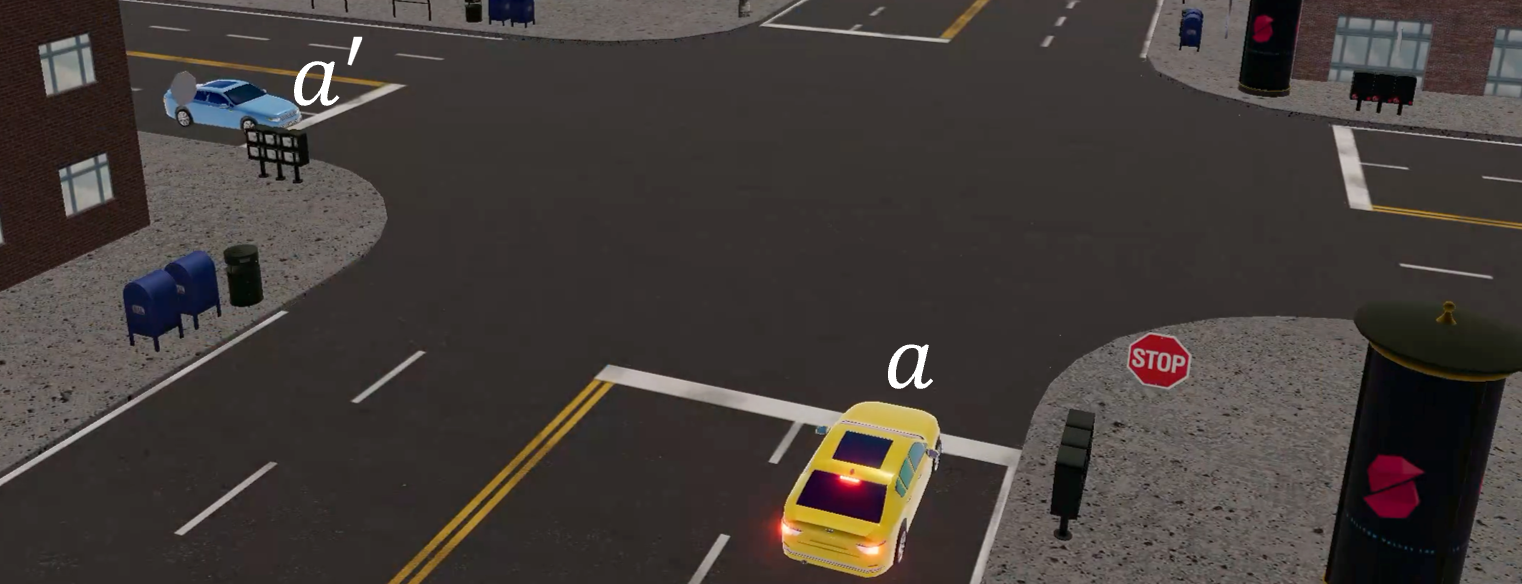}
			\end{minipage}%
		}%
   \subfigure[Violation of $p_2$ (at time $t$)\label{fig:simp2next}]{
			\begin{minipage}[t]{0.45\linewidth}
				\centering				\includegraphics[width=0.91\linewidth]{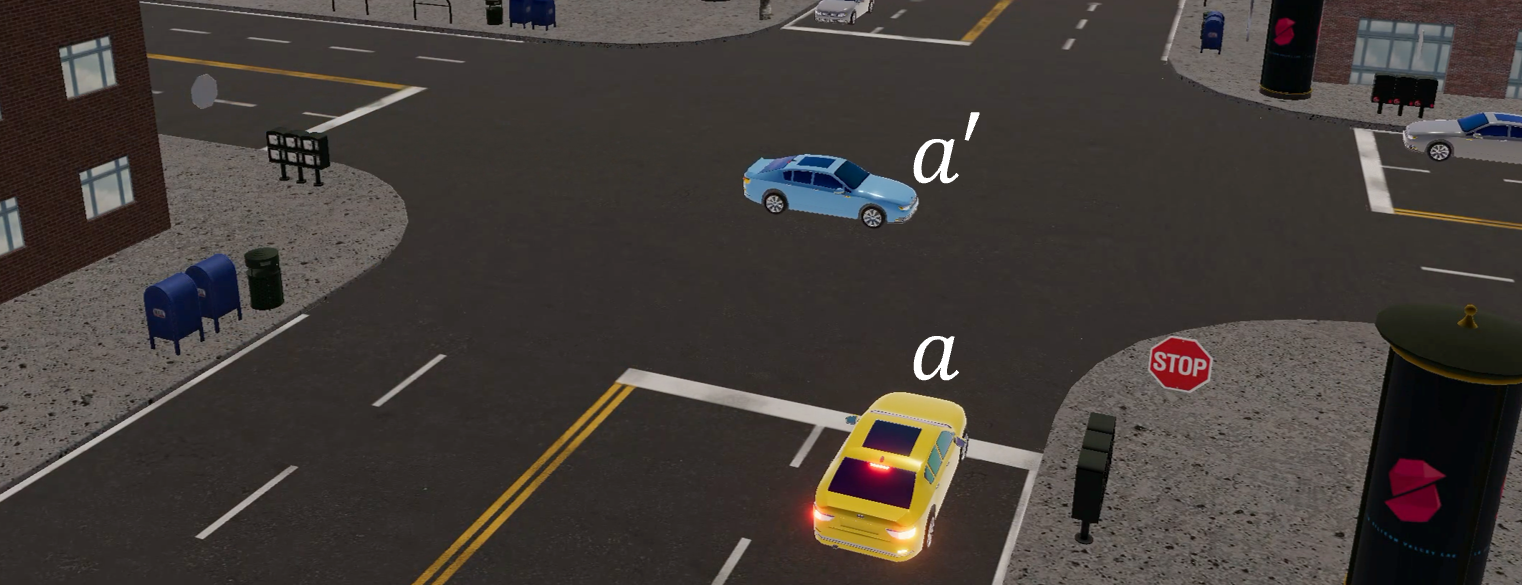}
			\end{minipage}%
		}%
   \\
   \subfigure[Violation of $p_5$ (at time $t-1$)\label{fig:simp5}]{
			\begin{minipage}[t]{0.45\linewidth}
				\centering				\includegraphics[width=0.91\linewidth]{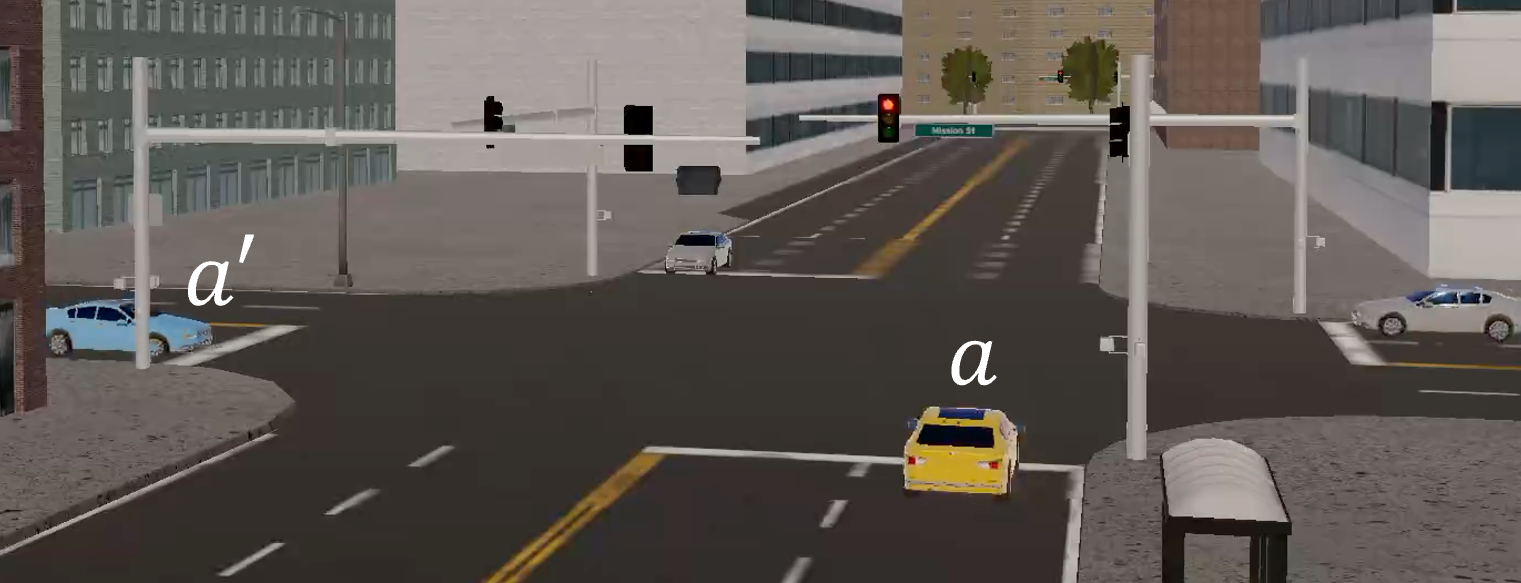}
			\end{minipage}%
		}%
   \subfigure[Violation of $p_5$ (at time $t$)\label{fig:simp5next}]{
			\begin{minipage}[t]{0.45\linewidth}
				\centering				\includegraphics[width=0.91\linewidth]{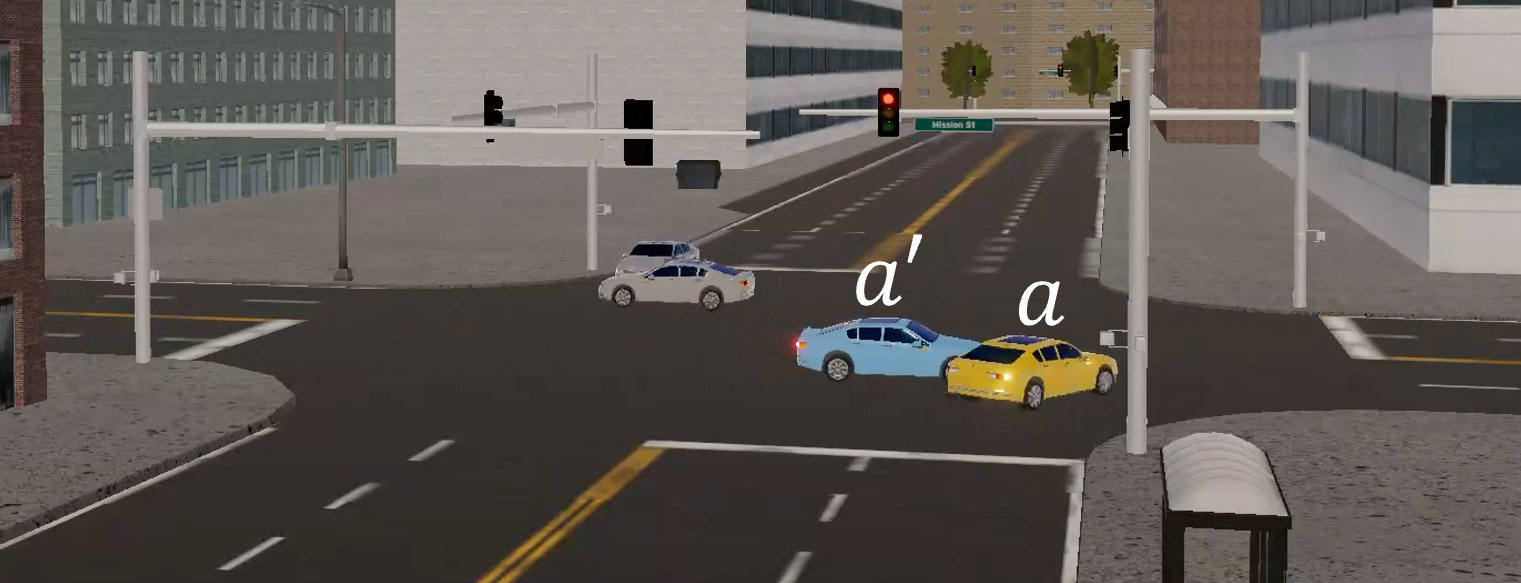}
			\end{minipage}%
		}%
   \hfill
   \subfigure[Violation of $p_6$ (at time $t-1$)\label{fig:simp6}]{
			\begin{minipage}[t]{0.45\linewidth}
				\centering				\includegraphics[width=0.91\linewidth]{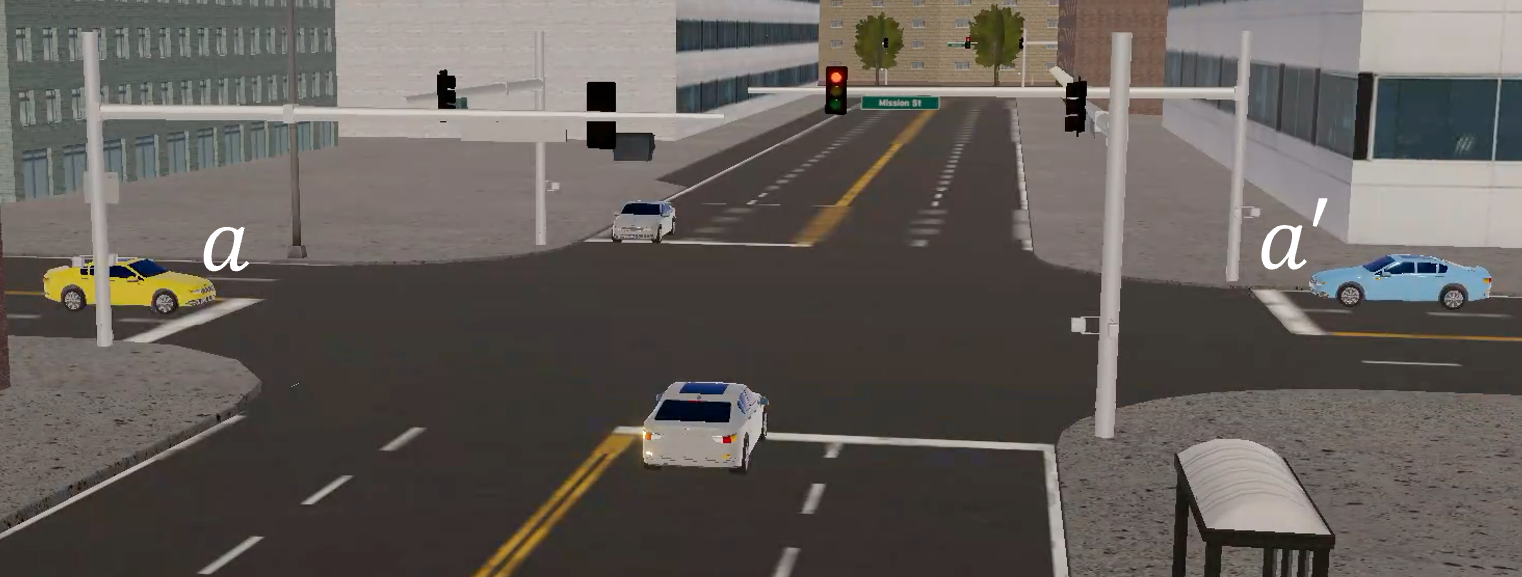}
			\end{minipage}%
		}%
   \subfigure[Violation of $p_6$ (at time $t$)\label{fig:simp6next}]{
			\begin{minipage}[t]{0.45\linewidth}
				\centering				\includegraphics[width=0.91\linewidth]{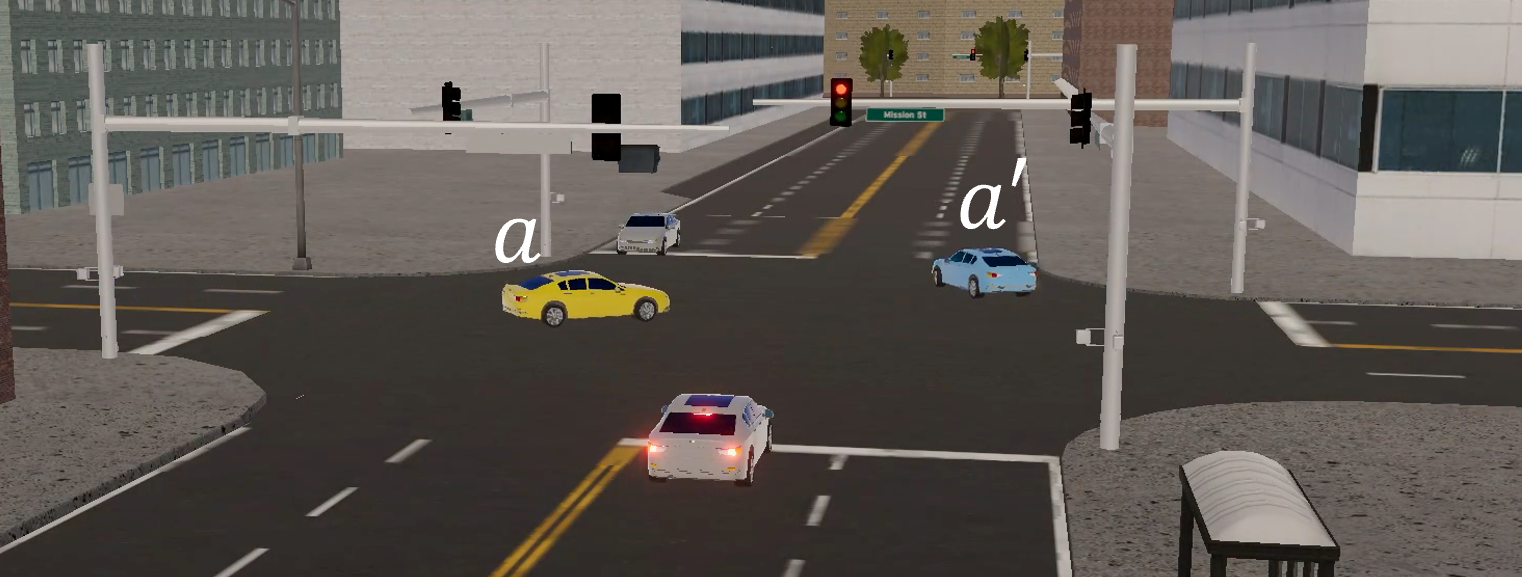}
			\end{minipage}%
		}
		
	\caption{Snapshots for property violations}
	\label{fig3} 
\end{figure*}

\vspace{15pt}
{\bf 1) Analysis for  property $p_1$}

Property $p_1$: If a vehicle is in the junction, then no other vehicle can be in the junction. Hence, violating $p_1$ means that more than one vehicle is in the junction at the same time.


This property is violated in all scenarios $(1, A), (2, A), (1, B)$ and $(2,B)$.

For scenarios $(1, A)$ and $(2, A)$, the cause is deficiency {\bf I$_1$} in Table \ref{table-issues}, due to the lack of controllability for short distances. During initialization, the controller randomly assigns a rate for vehicle acceleration without considering the safe braking distance. When the distance to go is 0.01 meters, the vehicle cannot brake safely.

For scenarios $(1, B)$ and $(2, B)$, the cause is  deficiency {\bf I$_2$} in Table \ref{table-issues}, due to hidden guidance for control. Vehicle control uses a boundary zone for junctions that is smaller than the area defined by its entrances and exits. For example, in Fig.~\ref{fig:issue1}, the shaded square area is considered as the junction area by the Scheduler. 
When the vehicle at entrance $0$ leaves the shaded square area, the vehicle at entrance $1$ is then scheduled to enter. 
In such a case, if we consider that the junction is delimited by its entrances and exits, the two vehicles are in the junction simultaneously.

We can also observe that even for the same class of scenarios, e.g., $(2, C)$, some scenario violates property $p_1$, while some do not. For example, there is one scenario in $(2,C)$ satisfying this property. 
The inconsistency is also due to the use of a boundary zone different from the area of the junction. 
Moreover, the boundary zone  is not placed in the center of the junction, or symmetrical under rotation. 
Thus, the validity of a property for equivalent scenarios may depend  on the configurations between vehicles and the boundary zone of the junction considered by the Scheduler.

For example, the first scenario shown in Fig.~\ref{fig:issue1} of $(2,C)$ does not satisfy the property $p_1$, for the same reason as discussed for scenarios $(1,B)$ and $(2,B)$. 
 However, when considering the case of Fig.~\ref{fig:issue2}, the third scenario in class $(2,C)$, 
 the vehicle at entrance $2$ enters the junction first. 
 As the boundary zone is close to exit $0$ and $1$, 
 the vehicle at entrance $3$ will enter the junction after the one at $2$ leaves the junction.
 The same is true for vehicles from entrances $3$ and $0$, and from entrances $0$ and $1$, respectively. 
 Therefore, there is no more than one vehicle in the junction simultaneously,
 and the property $p_1$ is satisfied.


{\bf 2) Analysis for property $p_2$}

Property $p_2$: If vehicles arrive at the stop sign at the same time, the one on the right has the right-of-way.


Property $p_2$ is violated in all scenarios. 
For scenarios $(1,A)$ and $(2,A)$, the cause of violation is explained by deficiency {\bf I$_1$} in Table \ref{table-issues}, as the vehicles cannot brake safely and enter the junction.

For scenarios $(1,B), (1,C), (1,D),(2,B), (2,C)$, and $(2,D)$, 
the cause of violation is explained by deficiency {\bf I$_3$} in Table \ref{table-issues}.
In these scenarios, the Scheduler ignores the priority rule. The four vehicles reach the stop sign at the same time. 
According to $p_2$, a circular priority relationship between the vehicles would result in a deadlock, as each of the four vehicles waits for the vehicle to its right to proceed.  
In the Simulator however, the Scheduler ignores the rule and schedules the vehicles according to the order of initialization.

For scenarios $(1,E)$ and $(2,E)$, 
the reason of violation is also explained by deficiency {\bf I$_3$}. Consider the first scenario in $(1,E)$ as an example. The vehicles at entrances $0$ and $1$ reach the stop sign of the junction at the same time. And according to the rule, the vehicle at entrance $1$ has a higher priority than the one at $0$. 
However, the Simulator schedules the vehicles according to the initialization order ignoring this rule.

Figs.~\ref{fig:simp2} and  \ref{fig:simp2next} show two successive scenes illustrating a violation. 
In Fig.~\ref{fig:simp2}, vehicles $a$ and $a'$ arrive at the same time. Vehicle $a'$ is to the right of $a$ and supposed to have higher priority. However, in the next moment, as shown in Fig.~\ref{fig:simp2next}, vehicle $a$ enters before $a'$.

{\bf 3) Analysis for property $p_3$}

Property $p_3$: The vehicle that arrives at the entrance first will proceed before the other vehicles. 

In the scenarios other than those with configuration $E$, all the vehicles arrive at the entrance simultaneously,
so the above property is not applicable and the corresponding results are labelled by NA. 
For the scenarios in $(1,E)$ and $(2,E)$, 
property $p_3$ is satisfied.
That is, the Simulator applies this first-in, first-out policy.

\subsubsection{Experimentation with a 4-way traffic light junction.}

The results of the validation of the properties of Table~\ref{tab:rules} for the four-way traffic light junction are shown in Table~\ref{tab:lt-eqnew}.  
We choose two structural equivalence classes different from those in Table~\ref{tab:eq4waystop}. 
The initial distances to the entrance and speeds of the vehicles are also chosen from $\{0.01m, 0.3 m, 20 m\}$ and $\{0 m/s, 10 m/s\}$, respectively. 
Experiments show that, for a traffic light junction, the vehicles follow for conflict resolution an implicit order encoded in the lanes of the junction. This is a source of inconsistencies as discussed below. 
%
%

{\bf 1) Analysis for property $p_4$}

Property $p_4$: Every vehicle facing a red light should stop until the traffic light turns green, unless the vehicle is turning right.

In scenarios $(3,F)$, the property is violated. The cause is deficiency {\bf I$_1$} in Table~\ref{table-issues}. The vehicle  facing red light cannot brake safely when it is making a left turn.  

In the other scenarios obtained from abstract scenarios of class  3, this property is satisfied, and all the vehicles wait before the traffic lights turns green.
However, for scenarios obtained from class 4, this property is not applicable since the two vehicles facing the red light are turning right.

{\bf 2) Analysis for property $p_5$}

Property $p_5$: If a vehicle facing a red light is turning right, then the vehicle should wait until there is no vehicle on its left. 

The abstract scenario 3 does not involve a vehicle turning right when facing a red light. Therefore, the property is not applicable for this class.

For scenarios in $(4, F)$ and $(4,G)$, the property is satisfied. 
In these scenarios, the vehicles facing green light will pass through the junction without decelerating. Then the vehicle facing the red light turns right.   


For scenarios in $(4,H)$ and $(4,I)$, the vehicles turning right at entrances $1$ and $3$ are facing red lights and will decelerate first. When they arrive at the entrances, the other vehicles have already entered the junction. Therefore, the property is not applicable.

The scenarios in $(4, J)$ do not satisfy property $p_5$. As shown in Fig.~\ref{fig:issue3}, the vehicle facing a red light can enter the junction when there is a vehicle at the entrance on its left. The reason described by deficiency {\bf I$_4$}  of Table \ref{table-issues} is that when a vehicle facing green light arrives at the junction, the  vehicle making a right turn may have already reached its randomly-assigned waiting time and enters the junction without considering the priority. 

We provide a snapshot for such a case in the simulation. In Fig.~\ref{fig:simp5}, vehicle $a$ is waiting to turn right facing a red light, and vehicle $a'$ arrives. In Fig.~\ref{fig:simp5next}, the traffic light for vehicle $a$ is still red. However, it enters the junction and nearly causes a collision. 
%

{\bf 3) Analysis for property $p_6$}

Property $p_6$: If two vehicles arrive at the entrances of a junction from opposite directions and the traffic lights are green, the vehicle turning left must give way to the other.

For class 3, property $p_6$ is not satisfied. 
For scenarios $(3,F)$, the reason is deficiency {\bf I$_1$} of Table~\ref{table-issues}, that is lack of controllability for short distances. The vehicle turning left cannot stop safely and enters the junction before the other. 

For the other scenarios from class 3, the reason is deficiency {\bf I$_5$} of Table~\ref{table-issues} due to the application of an incomplete priority order. For example, in Fig.~\ref{fig:issue4} the vehicle at entrance $0$ is turning right, and the vehicle at entrance $2$ enters without considering that the one at entrance $0$ has a higher priority.
However, in the abstract scenario 4, no vehicle turns left. Thus this property is not applicable. 

We provide snapshots showing the violation of the property in  Figs.~\ref{fig:simp6} and~ \ref{fig:simp6next}. In Fig.~\ref{fig:simp6}, vehicles $a,a'$ arrive at the entrances simultaneously, where $a$ is turning left and $a'$ is turning right. According to the traffic rule, $a'$ should proceed first. However, in Fig.~\ref{fig:simp6next}, the two vehicles enter simultaneously. 

\section{Related work}

In the literature of simulation-based validation for 
autonomous driving systems~\cite{koopman2016challenges,lou2022testing}, there is a large body of work on the generation of 
safety-critical scenarios, either by using scenario modeling languages~\cite{majumdar2021paracosm,fremont2022scenic}, 
or from  available databases~\cite{TianJWY00LY22}. 

Scenic is a well-known probabilistic programming language for generating  scenarios for autonomous driving systems~\cite{fremont2022scenic}. 
Paracosm~\cite{majumdar2021paracosm} is another software system that also allows users to  describe complex driving situations with specific characteristics and generate scenarios by different parameter configurations. 
Our approach is distinguished by its rigor and completeness because it is based on a semantic model extracted from the Simulator. We generate scenarios based on coverage criteria and test them against properties specified in a logic and verified by verification techniques.  

Many works adopt search-based algorithms to generate scenarios challenging the autonomous driving systems. For example, Av-fuzzer~\cite{2020AV-fuzzer} utilizes a genetic algorithm-based search to
 detect situations where an autonomous driving system can run into safety violations.
 The basic idea of the search is to perturb the driving maneuvers of traffic participants 
 and optimize the safety constraints based on vehicle dynamics. 
MOSAT~\cite{TianJWY00LY22} uses genes to encode basic driving maneuvers and applies a multi-objective genetic algorithm to search for adversarial and diverse test scenarios. None of them considers violation of properties or traffic rules as the criteria in identifying safety-critical scenarios.

A recent work \cite{Zhong2022GuidedCD} proposes a controllable traffic generation method based on broadcast modeling and signal temporal logic (STL). 
The proposed method allows the generation of realistic traffic trajectories that mimic the real traffic  and meet the desired objectives defined by the STL formulas. However, it does not address issues of validation or verification. It is still unclear 
whether the generated traffic trajectories are effective and efficient in revealing
potential safety issues for ADS.

Efforts are also made to generate scenarios according to map topology. 
To facilitate the virtual testing of motion planners for automated vehicles,
 the work in  \cite{Klischat2020c} first extracts a large variety of road networks from OpenStreetMap. 
 Then it uses the traffic simulator SUMO to generate traffic scenarios for these road networks. The criticality of the scenarios is reinforced by the use of nonlinear optimization. 
The work in \cite{TangZ0WLW21} classifies junction lanes according to the collision avoidance maneuvers of the ego vehicle and builds map topology-based scenarios with a genetic algorithm. SOCA~\cite{butz2020soca} adopts zone graphs as the abstraction of traffic situations at junctions for behavior analysis, where each zone graph represents an intention of the ego vehicle. 
Unlike these two works~\cite{TangZ0WLW21,butz2020soca},  we ignore the detailed topology of the junctions while we consider abstractions that suffice to define coverage criteria for a systematic exploration of traffic patterns. 

In addition to generating scenarios, many works use linear temporal logic~\cite{esterle2020formalizing,Gressenbuch2021,sifakis-framework}, or STL~\cite{Zhong2022GuidedCD} to describe traffic rules or safety properties of  autonomous driving systems.   We also adopt
linear temporal logic to describe traffic rules. However formulas are parametric involving quantification over  both map and vehicle attributes. 

Testing techniques have also been applied for the validation of ADS,
such as combinatorial testing~\cite{majumdar2021paracosm,comopt2022},
metamorphic testing~\cite{zhang2018deeproad,zhou2019metamorphic} 
and fuzz testing~\cite{2020AV-fuzzer}.
Among these techniques, adopting combinatorial testing can guarantee  the coverage of the generated scenarios with respect to the given parameters~\cite{majumdar2021paracosm,comopt2022}.  
By adopting metamorphic relations on the inputs, such as generating scenes with various weather conditions~\cite{zhang2018deeproad}, or introducing noise~\cite{zhou2019metamorphic}, 
the inconsistency between the outputs can be detected, which eliminates the need to use a test oracle. 
In fuzz testing, existing scenarios can be mutated to search for potential bugs or safety violations~\cite{2020AV-fuzzer}.  
In our work, we use equivalence on scenarios as a metamorphic testing relation. 
However, we use an Oracle applying runtime verification to detect discrepancies.

\section{Conclusion}


The paper proposes a rigorous simulation-based method for the validation of autonomous driving systems. A key idea is to link the LGSVL Simulator with the RvADS tool that combines a Scenario Generator and a Monitor to test sets of desired system properties specified in a linear temporal logic. The Scenario Generator can control the execution of the validated system to reach global states involving a high risk of violating the tested properties. The Monitor applies runtime verification to check the runs resulting from the application of the scenarios against the properties.

The integration of LGSVL in RvADS required considerable engineering effort in two directions. On the one hand, the modification of the simulation runtime by adding a Controller that enforces the application of scenarios. For a given set of vehicles, a scenario is a set of itineraries with given initial conditions, one for each vehicle. On the other hand, the application of runtime verification required the development of a specific API to export a semantic model of the tested system. The latter includes a metric graph representing the  system's static environment and extracted from a HD map of the Simulator. 

Unlike simulation-based approaches that focus on quantitative criteria such as the number of hours or kilometers simulated, we rely on the semantic model of the system to achieve sufficient coverage of high-risk situations. We focus on testing traffic rules applicable to two types of junctions where the probability of accidents is increased. The main result is the definition of a property-preserving equivalence on scenarios. We show how we can use equivalent scenarios to discover flaws in the simulated system and ensure fair coverage of risky situations whose probability of occurring in random simulation is very low.

The obtained experimental results reveal several deficiencies in the simulated systems,
some of which are related to the driving policies of the agents and some to the simulation runtime implementation. 
In addition, we found that the deceleration rates applied by the simulated vehicles are unreasonable. For example, the speed of a vehicle in the simulator can be reduced from 10 m/s to 0 m/s in less than 3.5 meters, which is impossible in reality. 

The method is novel to our knowledge. It shows that for this reputedly difficult problem, there is a way to tame its complexity. Instead of exploring ordinary situations with low risk potential over long periods of time, we can decompose the validation problem into relatively independent contexts given the locality of the decision making. In a future work, we will show how to apply a compositionality principle to achieve global system validation.
\bibliographystyle{ACM-Reference-Format}
\bibliography{reference}

\end{document}